\documentclass[12pt]{emulateapj}
\usepackage{graphicx}

\shorttitle{High Signal-to-Noise ULX Study}
\shortauthors{Winter \& Mushotzky}

\begin{document}
\title{Elemental Abundances of Nearby Galaxies through High Signal-to-Noise {\it XMM-Newton} Observations of ULXs}

\author{Lisa M. Winter}
\affil{ Astronomy Department, University of Maryland, College Park, MD 20742}
\email{lwinter@astro.umd.edu}

\author{Richard F. Mushotzky}
\affil{Goddard Space Flight Center, Greenbelt, MD 20771}
\email{richard@milkyway.gsfc.nasa.gov}

\author{Christopher S. Reynolds}
\affil{ Astronomy Department, University of Maryland, College Park, MD 20742}
\email{chris@astro.umd.edu}

\begin{abstract}
In this paper, we examined {\it XMM Newton} EPIC spectra of 14 ultra-luminous
X-ray sources (ULXs)
in addition to the {\it XMM} RGS spectra of two sources (Holmberg II X-1 and Holmberg IX X-1).
We determined oxygen and iron abundances of the host galaxy's interstellar medium (ISM) using
K-shell (O) and L-shell (Fe) X-ray photo-ionization edges towards these ULXs. 
We found that the oxygen abundances closely matched recent solar abundances for
all of our sources, implying that ULXs live in similar local environments despite the wide range
of galaxy host properties.  
Further, the ISM elemental abundances of the host galaxies, as indicated from the O/H values 
obtained from the X-ray spectral fits, are in good agreement with the 
O/H values obtained by Garnett (2002) from studies of \ion{H}{2} regions in the 
same external galaxies.  We find roughly solar O/H values independent of the
host galaxy luminosity, star formation rate, or the ULX location in the galaxy.
Also, we compare the X-ray hydrogen column densities (n$_H$) for 8 ULX
sources with column densities obtained from radio \ion{H}{1} observations. 
The X-ray model n$_H$ values are in good agreement with the \ion{H}{1} n$_H$ values, 
implying that the hydrogen absorption towards the ULXs is not local to the source
(with the exception of the source M81 XMM1).   In order to obtain the column density and
abundance values, we fit the X-ray spectra of the ULXs with a combined power law and
one of several accretion disk models.  We tested the abundances obtained from the 
XSPEC models {\tt bbody}, {\tt diskbb}, {\tt grad}, and {\tt diskpn} along with a power law, finding
that the abundances were independent of the thermal model used.  We comment on the
physical implications of these different model fits.  We also note that very deep observations
allow a breaking of the degeneracy noted by \citet{sto06} favoring a high mass solution for
the absorbed {\tt grad} + power law model. 
    
\end{abstract}

\section{Introduction}

Long exposure XMM-Newton observations of nearby galaxies offer new opportunities to study various properties of ultraluminous 
X-ray source (ULX) spectra.  ULXs are bright, non-nuclear X-ray sources with X-ray
luminosities $> 3 \times 10^{39}$\,erg\,s$^{-1}$.  In a previous paper \citep{win05} (Paper~1)\defcitealias{win05}{Paper~1}, 
we analyzed ULX spectra from 32 nearby 
galaxies.  Based on spectral form, luminosity, and location within the optical host galaxy, we
classified a population of high/soft state and low/hard state ULXs.  For the ULX sources with the 
greatest number of counts, the high signal-to-noise {\it XMM} observations can be analyzed with
more than the simple schematic models of our first study (an absorbed power law 
and blackbody model for the high-state
and an absorbed power law model for the low state).  Particularly, the spectra can be used
to investigate the properties of absorption along the line of sight to ULX sources.

Many similar studies have been done within our own Milky Way, where
X-ray absorption models have been used to determine
column densities and abundances of the interstellar medium (ISM).  The procedure
involves spectral fits of absorption features in bright, background X-ray sources.
Successful determinations have been made using background galaxy clusters \citep{bau05}
and X-ray binaries \citep{jue04}.  These studies made use of a bright, X-ray
source as a background through which they can observe the 542\,eV absorption edge
produced by photo-ionization of the inner K-shell electrons of oxygen.  Analogous studies
have been used in the radio (see \citet{dic90}) to optical regime, using 
quasars, supernovae, or stars as a background for hydrogen
absorption and 21-cm emission, as a means to measure hydrogen column densities and
metal abundances.

In this study, we extend the X-ray absorption studies to external galaxies using
ultra luminous X-ray sources.  Due to their extreme brightness
in the X-ray regime and their non-nuclear location in external galaxies, these sources are ideal
for probing the ISM of their host galaxies.  Typical ULXs, from our \citetalias{win05} study,
have Galactic line-of-sight column densities of a few $10^{20}$\,cm$^{-2}$
\citep{dic90} and measured X-ray column densities greater than $10^{21}$\,cm$^{-2}$ 
(Fig. 8 of \citetalias{win05}) for the combined ULX and host galaxy.  Thus, if the local environment
of the ULX contributes little absorption, the X-ray column density is dominated by the host
galaxy.  One goal of this study is to determine whether this absorption is that of the host galaxy
or local ULX environment.  Therefore, we compare the X-ray measured hydrogen column density
with \ion{H}{1} measurements from alternate methods.
 
In addition to the brightness of ULXs and the relatively small Milky Way contribution to their
X-ray hydrogen column densities, their well characterized X-ray spectra make ULXs ideal
for measuring absorption features of the ISM.  Bright ULXs (e.g. NGC 1313 X-1) typically have 
spectra that are well-fit by an absorbed multi-component blackbody and power law model. 
However, there is discussion over whether this standard model is the most physical model
for the ULXs (see, for example, \citet{sto06} or \citet{gon06}).  
Different models applied to the base ULX spectra can affect the  
absorption measurements, particularly in the softer part of the spectrum.  Thus, in this paper
we investigate the effect different soft component models have on the X-ray
measured hydrogen column density and elemental abundances (through the oxygen K-shell edge
at 542\,eV and the iron L-shell edge at 851\,eV).

We use high signal-to-noise {\it XMM Newton} observations of ULXs to measure
hydrogen column densities and elemental abundances of oxygen and iron.  
Located in external galaxies, the X-ray spectral resolution of available ULX spectra is not as good as those of
Galactic X-ray binaries, which often have grating spectra available (e.g. \citet{jue04}).
Therefore, in order to be
able to distinguish the oxygen K-shell edge as well as the iron L-shell edge located
at 851\,eV, we needed observations with a large number of counts ($\approx 5000$\,counts).
  The X-ray observatory {\it XMM-Newton},
having a larger collecting area than Chandra, provides the counts necessary in order
to conduct this study.  Further, with recent 100\,ks {\it XMM-Newton} observations
available for the host galaxies of two well-studied ULXs (Holmberg II and Holmberg IX), these observations allow for the added analysis of Reflection Grating Spectrometer 
(RGS) spectra in addition to
spectra from the European Photon Imaging Cameras (EPIC).  The spectral resolution of the
EPIC and RGS allow us to test different soft component models for the ULX sources, to determine
the effect of the model on absorption and abundance measurements.

\section{Source Selection and Data Reduction}

In \citetalias{win05} we conducted an archival {\it XMM-Newton} study of ULXs in 32 nearby
($< 8$ Mpc) galaxies.  As described in that paper, we extracted spectra for the brightest
sources in the observations, corresponding to $> 400$\,counts.
In this study, we chose to further analyze the spectra of the objects with
the highest number of counts ($> 5000$\,counts\footnotemark). In addition to the 11 sources from \citetalias{win05}, 
we include an analysis of 3 additional sources: the two ULXs in the 
spiral galaxy 
NGC 4559 (observation 0152170501) and
the bright source in M33 (observation 0102640101).  A full list of the 14 ULX sources, with details of the observations (including exposure
times and count rates), is found in Table~\ref{tbl-5}.   

\begin{deluxetable*}{lllllll}
\tablecaption{Details of the ULX Sources Analyzed\label{tbl-5}}
\tablewidth{0pt}
\tablehead{
\colhead{Source} & \colhead{RA (h m s)\tablenotemark{a}} 
& \colhead{Dec ($\circ\ \prime\ \prime\prime$)\tablenotemark{a}} & \colhead{n$_{H GAL}$\tablenotemark{b}} 
& \colhead{Obs ID} & \colhead{Exposure Time (s)\tablenotemark{c}} & \colhead{Count Rate (cts\,s$^{-1}$)\tablenotemark{c}}
}
\startdata
NGC247 XMM1 & 00 47 03.8  & -20 47 46.2 & 1.54 & 0110990301& 3458, 1389, 1379    & 0.20, 0.06, 0.06 \\
NGC253 XMM2 & 00 47 22.4  & -25 20 55.2 & 1.40 & 0152020101&  -, 10347, 10304       & -, 0.08, 0.09 \\
NGC300 XMM1 & 00 55 09.9  & -37 42 13.9 & 3.11 & 0112800101& 6778, 2248, 2453     & 0.19, 0.05, 0.05 \\
M33 X-8               & 01 33 50.9  & +30 39 36.1& 5.58& 0102640101& 11919, 12147, 12142& 0.06, 0.02, 0.02 \\
NGC1313 XMM3& 03 18 22.5 & -66 36 06.2 & 4.0  & 0106860101 & 6960, 2179, 1793     & 0.233, 0.08, 0.07 \\
Holm II XMM1     & 08 19 28.8  & +70 42 20.3& 3.42& 0112520701 & 31052, 1257, 10807& 2.72, 0.76, 0.73 \\
\nodata                 & \nodata       & \nodata    & \nodata& 0200470101& 56987,65758, 65766  & 3.29, 0.95, 0.94 \\
\nodata                 & \nodata       & \nodata    & \nodata& 0200470101 & 41816, 41802 (RGS) & 0.09, 0.11 (RGS) \\
M81 XMM1          & 09 55 32.9  & +69 00 34.8& 4.12& 0111800101 & 50788, -, 18988        & 0.51, -, 0.22 \\
\nodata		   & \nodata       & \nodata &\nodata    & 0200980101 & -, 111910, 111930  & -, 0.111, 0.104 \\
Hol IX XMM1       & 09 57 53.3  & +69 03 48.7& 4.0   & 0112521001 &14976, 6546, 6586   & 2.07, 0.64, 0.65 \\
\nodata		   & \nodata       & \nodata &\nodata    & 0200980101 & 104010, 111760, 111830& 1.69, 0.51, 0.51 \\ 
\nodata		   & \nodata       & \nodata &\nodata    & 0200980101 & 103840, 103830 (RGS) & 0.03, 0.04 (RGS) \\
NGC 4559 X7	   & 12 35 51.8  & +27 56 04   & 1.51 & 0152170501 & 34517, -, -                  & 0.318, -, - \\
NGC 4559 X10   & 12 35 58.6  & +27 57 40.8& 1.51 & 0152170501 & 34509, -, -                  & 0.238, -, - \\
NGC4631 XMM1& 12 41 55.8  & +32 32 14   & 1.28 & 0110900201 & 5093, 1969, 1762    & 0.13, 0.04, 0.04   \\
NGC5204 XMM1& 13 29 38.5  & +58 25 03.6& 1.42 & 0142770101 & 9981, 3352, 3384    & 0.628, 0.177, 0.179 \\
\nodata                 & \nodata        & \nodata   & \nodata& 0142770301 & 9231, 2284, 2349     & 0.855, 0.247, 0.258 \\
M83 XMM1          & 13 37 19.8  & -29 53 49.8 & 3.94 & 0110910201 &3074, 927, 987           & 0.12, 0.033, 0.025 \\
NGC5408 XMM1\tablenotemark{d}& 14 03 19.8  & -41 22 59.3 & 5.73 & 0112290601 & 5932, 2036, 2077     & 0.128, 0.032, 0.033 \\
\enddata
\tablenotetext{a}{RA and Dec values quoted are the source positions from the EPIC-pn images.}
\tablenotetext{b}{Milky Way hydrogen column density along the line of sight from
\citet{dic90} in units of $10^{20}$\,cm$^{-2}$}
\tablenotetext{c}{Exposure times and count rates are listed for the EPIC pn, MOS1, MOS2, and RGS (1 and 2), where available. Note that for sources with multiple observations, details for the additional observations are listed below the first observation.  Details of the specific RGS (RGS1 and RGS2) observations are indicated with (RGS).}
\tablenotetext{d}{A second observation of NGC 5408 XMM1 is referred to in Section 4.2.  This
observation (0302900101) with an exposure time of 130335\,s is proprietary 
and an analysis will appear in Strohmayer {\it et al.} (in prep).}
\end{deluxetable*}

\footnotetext{In the appendix we show results of simulations to determine the number of
counts necessary to detect the oxygen and iron absorption edges.  We found that 
$> 40000$\,counts are needed to constrain iron without having large errors in the measurement.
For oxygen, $> 5000$\,counts are needed to constrain oxygen without large errors.  See the
appendix for further details.}

The two ULX sources in NGC 4559 were originally studied by \citet{vog97} using ROSAT and \citet{cro04} with 
{\it XMM-Newton}.  We follow the naming convention established in these papers. Both of these
sources (X7 and X10) were not included in \citetalias{win05} because the host galaxy's distance is greater than the 8\,Mpc limit we initially
required.  However, we include these sources now due to the high number of counts
($> 5000$\,counts) in their spectra.  Initially we did not include M33 X-8 in our ULX survey due to its location in the center of its host galaxy.  This source, 
however,
shows no evidence of being a low-luminosity AGN and is more likely a black hole X-ray binary \citep{tak94}.

Since the initial study of \citetalias{win05}, longer exposure time {\it XMM-Newton} observations have become available for three of our 
sources from the original study.
With the permission of Tod Strohmayer, we include data from the 100\,ks {\it XMM-Newton} observation (0200980101) of 
Holmberg IX.
These data include pn and MOS spectra of Holmberg IX XMM1 and MOS spectra of M81 XMM1.  We also include a
100\,ks {\it XMM-Newton} observation (0200470101) of Holmberg II XMM1 that became public after the \citetalias{win05} study.   
The EPIC and RGS spectra from this observation of Holmberg II XMM1 were first analyzed by \citet{goa05}.
For Holmberg IX XMM1 and Holmberg II XMM1, the 100\,ks exposures provided us with the opportunity to extract and analyze, in addition to the EPIC spectra, spectra from 
the RGS detectors.  Thus, we include an analysis of RGS spectra for both Holmberg II XMM1 and Holmberg IX XMM1.


For the EPIC spectra we added to our original sample, we followed the same reduction 
method as in \citetalias{win05}.  
For observations that were processed with an earlier version of 
the {\it XMM-Newton} Science Analysis System (SAS) (we used SAS version 6.0\footnotemark), the
observation data files (ODF) were used to produce calibrated photon event
files for the EPIC-MOS and pn cameras using the commands {\tt emchain}
and {\tt epchain}.   The events tables were filtered
using the standard criteria outlined in the {\it XMM ABC Guide}.  For
the MOS data (both MOS1 and MOS2 cameras), good events constitute
those with a pulse height in the range of 0.2 to 12 keV and event
patterns that are characterized as 0-12 (single, double, triple, and
quadruple pixel events).  For the pn camera, only patterns of 0-4
(single and double pixel events) are kept, with the energy range for
the pulse height set between 0.2 and 15 keV.  The selection
expression ``FLAG == 0'' was used to exclude bad pixels and events
too close to the edges of the CCD chips.  Time filtering was applied as needed
by editing the light curve produced in {\tt xmmselect}.  For the EPIC observations, 
time periods in the observation with high count rates (flares) in the MOS and
pn were 
 cut using the command {\tt tabgtigen} with the `RATE$<$' command set to
 5\,cts\,s$^{-1}$ for MOS detectors and 20\,cts\,s$^{-1}$ for the pn detector.

We extracted source and background spectra along
with response and ancillary response matrices using the SAS task {\tt especget}.
The source spectra were extracted from circular regions, typically with radii of 20\,arcseconds.
This region was adjusted depending on the size of the source and the proximity of the
source to either another source or the edge of a CCD chip.
We extracted background spectra from annular regions, except when the source was
near another source or near the edge of a chip.  In this case, in order to avoid source 
confusion, we extracted background spectra from a 
circular region located near the source and on the 
same CCD chip (as for Holmberg IX XMM1 
and M81 XMM1).

For the RGS spectra we extracted first and second order spectra for the sources 
Holmberg IX XMM1 (0200980101) and
Holmberg II XMM1 (0200470101) using the {\tt rgsproc} command. The RA and Dec values used
to extract the RGS spectra were obtained from the EPIC pn data and are the values quoted
in \citetalias{win05}.  Time filtering was applied as for the EPIC data with {\tt tabgtigen}, where
 the `RATE$<$' command was set to 
0.5\,cts\,s$^{-1}$ for Holmberg IX and 0.1\,cts\,s$^{-1}$ for Holmberg II.  Once the
spectra were obtained, for the RGS as well as EPIC data, they were rebinned to require at least 20
counts per bin, using the command {\tt grppha} in LHEASOFT.


\footnotetext{Since the original processing of the data in this study, a new version
of SAS (6.5) had become available.  This version includes updates to the low energy
response files.  In the appendix, we show results of fitting the pn observations processed
with SAS 6.5 (for pn counts
$> 5000$).  We found that the model parameters are not significantly different
between the SAS 6.0 EPIC spectra and SAS 6.5 pn spectra. }

\section{Spectral Fitting}

Spectral fitting proceeded using XSPEC v11.3.1.  For the RGS spectra, we simultaneously fit the first 
order spectra from both RGS1 and RGS2 in the RGS band (0.33 - 2.5\,keV). 
For the EPIC spectra, we fit the pn and MOS spectra simultaneously in the
0.3-10\,keV energy range.  We allowed a free normalization constant to account
for the differences in flux calibration between the three EPIC cameras.  Both the RGS and
EPIC spectra were fit separately.

In the \citetalias{win05} study, we had fit all of the sources with three standard models:
an absorbed power law, an absorbed bremsstrahlung model, and an absorbed combined
blackbody and power law model.  We used the XSPEC model {\tt wabs} to account for
absorption from the Milky Way and the host galaxy/ULX contribution.  
This model is a photo-electric absorption 
model using the cross-sections of \citet{mor83} and the solar abundances of \citet{and82}. 
We found that the spectra of the brightest ULXs were typically best-fit by an absorbed
combined blackbody and power law model (fit in XSPEC as {\tt wabs*wabs*(bbody + pow)} where
the first {\tt wabs} model was fit to the \citet{dic90} Milky Way value and the second {\tt wabs} model
was fit to the remaining host galaxy/ULX contribution).  Likewise, we began our study by fitting the
additional sources (M33 X-8, NGC 4559 X-7, and NGC 4559 X-10) with the same three models
noted above.  We found that these sources were well fit by the absorbed blackbody and power law 
model with $\chi^2$/dof $\approx 1.0$. 

\begin{deluxetable*}{lllllllll}
\tablecaption{Spectral Fits for EPIC spectra with {\tt tbabs*tbvarabs*edge*(grad + pow)} model\label{tbl-2}}
\tablewidth{0pt}
\tablehead{
\colhead{Source} & \colhead{n$_{H}$\tablenotemark{a}} & \colhead{O abund.\tablenotemark{b}} 
& \colhead{Fe abund.\tablenotemark{b}} & \colhead{Mass (M$_{\sun}$)} 
& \colhead{$\dot{M}/\dot{M}_{Edd}$\tablenotemark{c}} & \colhead{$\Gamma$} & \colhead{$\chi^{2}$/dof} 
& \colhead{counts\tablenotemark{d}}
}
\startdata
HolmII XMM1& 0.12$^{+0.01}_{-0.01}$ & 0.90$^{+0.11}_{-0.11}$ & 0.0$^{+0.15}_{-0.0}$ &  141$^{+21}_{-32}$ & 0.15$^{+0.02}_{-0.09}$ & 2.36$^{+0.05}_{-0.05}$ &1.11 & 342874\\
\\
\nodata& 0.17$^{+0.02}_{-0.04}$ & 1.27$^{+0.23}_{-0.42}$ & 3.15$^{+1.29}_{-1.74}$ & 791$^{+209}_{-363}$ & 0.07$^{+0.03}_{-0.03}$ & 2.29$^{+0.07}_{-0.10}$ & 0.98 & 43116\\
\\
HolmIX XMM1& 0.19$^{+0.02}_{-0.02}$ & 1.34$^{+0.12}_{-0.13}$ & 2.07$^{+0.12}_{-0.13}$ & 382$^{+93}_{-76}$ & 0.09$^{+0.02}_{-0.02}$ & 1.38$^{+.025}_{-.025}$ & 1.05 & 148061\\
\\
\nodata& 0.33$^{+0.08}_{-0.09}$ & 1.38$^{+0.29}_{-0.40}$ & 3.91$^{+1.09}_{-1.86}$ & 1181$^{+1516}_{-719}$ & 0.11$^{+0.20}_{-0.07}$ & 1.73$^{+0.09}_{-0.09}$ & 0.64 & 28108\\
\\
M33 X-8& 0.20$^{+0.04}_{-0.04}$ & 1.04$^{+0.13}_{-0.18}$ & 1.74$^{+0.62}_{-0.74}$ & 5.24$^{+0.38}_{-0.35}$ & 1.02$^{+0.09}_{-0.08}$ & 2.83$^{+0.34}_{-0.31}$ & 0.97 & 123903\\
\\
M81 XMM1& 0.66$^{+0.11}_{-0.13}$ & 1.20$^{+0.08}_{-0.10}$ & 2.04$^{+0.44}_{-0.57}$ & 8.61$^{+33.58}_{-5.03}$ & 2.32$^{+44.8}_{-1.31}$ & 4.84$^{+0.58}_{-0.66}$ & 1.02 & 69776\\
\\
\nodata& 0.27$^{+0.21}_{-0.09}$ & 1.42$^{+0.58}_{-0.50}$ & 1.67$^{+1.61}_{-1.67}$ & 4.2$^{+7.4}_{-1.8}$ & 8.31$^{+4.36}_{-5.98}$ & 2.08$^{+1.58}_{-0.75}$ & 0.91 & 31731\\
\\
NGC253 XMM2 & 0.53$^{+0.07}_{-0.12}$ & 1.55$^{+0.20}_{-0.29}$ & 2.73$^{+1.17}_{-1.46}$ & 4140$^{+860}_{-2213}$ & 0.05$^{+0.03}_{-0.03}$ & 2.32$^{+0.10}_{-0.12}$ & 0.96 & 20651\\
\\
NGC5204 XMM1& 0.10$^{+0.04}_{-0.03}$ & 1.42$^{+0.55}_{-0.75}$ & 0.0$^{+1.84}_{-0.0}$ & 464$^{+366}_{-219}$ & 0.04$^{+0.08}_{-0.02}$ & 1.92$^{+0.08}_{-0.08}$ & 0.96 & 16717\\
\\
\nodata & 0.13$^{+0.05}_{-0.03}$ & 0.77$^{+0.60}_{-0.77}$ & 0.0$^{+1.70}_{-0.0}$ & 449$^{+490}_{-217}$ & 0.06$^{+0.12}_{-0.03}$ & 2.02$^{+0.13}_{-0.14}$ & 0.93 & 13864\\
\\
NGC1313 XMM3 & 0.67$^{+0.02}_{-0.04}$ & 1.37$^{+0.13}_{-0.14}$ & 0.0$^{+0.34}_{-0.0}$ & 5000$^{+0.03}_{-1160}$ & 0.08$^{+0}_{-0.02}$ & 2.66$^{+0.08}_{-0.09}$ & 1.02 & 10932\\
\\
NGC300 XMM1 & 0.15$^{+0.05}_{-0.04}$ & 2.46$^{+0.45}_{-0.43}$ & 0.0$^{+1.77}_{-0.0}$ & 417$^{+312}_{-175}$ & 0.02$^{+0.03}_{-0.01}$ & 2.46$^{+0.10}_{-0.11}$ & 1.01 & 11479\\
\\
N4559 X-7 & 0.16$^{+0.05}_{-0.03}$ & 0.46$^{+0.53}_{-0.46}$ & 0.0$^{+1.00}_{-0.0}$ & 755$^{+901}_{-443}$ & 0.06$^{+0.12}_{-0.02}$ & 2.10$^{+0.10}_{-0.10}$ & 0.83 & 12506\\
\\
NGC4631 XMM1 & 0.28$^{+0.06}_{-0.06}$ & 0.62$^{+0.43}_{-0.58}$ & 0.11$^{+1.86}_{-0.11}$ & 5.5$^{+11.4}_{-18.6}$ & 0.84$^{+0.88}_{-0.42}$ & 5.71$^{+2.03}_{-0.31}$ & 1.06 & 8824\\
\\
NGC5408 X-1 & 0.09$^{+0.03}_{-0.03}$ & 1.99$^{+0.26}_{-0.52}$ & 5.0$^{+0.0}_{-1.68}$ & 1477$^{+342}_{-477}$ & 0.10$^{+0.15}_{-0.09}$ & 2.53$^{+0.20}_{-0.16}$ & 0.96 & 10045\\
\\
NGC4559 X-10 & 0.12$^{+0.05}_{-0.04}$ & 1.28$^{+0.61}_{-0.81}$ & 0.0$^{+2.30}_{-0.0}$ & 5.34$^{+7.88}_{-2.75}$ & 2.74$^{+2.52}_{-1.86}$ & 2.09$^{+0.52}_{-0.28}$ & 0.77 & 8837\\
\\
NGC247 XMM1 & 0.44$^{+0.18}_{-0.12}$ & 1.09$^{+0.35}_{-0.28}$ & 0.0$^{+1.45}_{-0.0}$ & 1717$^{+3205}_{-929}$ & 0.05$^{+0.02}_{-0.03}$ & 3.60$^{+6.40}_{-6.60}$ & 0.60 & 6226\\
\\
M83 XMM1 & 0.13$^{+0.09}_{-0.10}$ & 1.70$^{+1.12}_{-1.65}$ & 0.0$^{+3.92}_{-0.0}$ & 6.34$^{+13.4}_{-4.3}$ & 0.92$^{+0.70}_{-0.70}$ & 2.64$^{+0.88}_{-0.64}$ & 0.83 & 4988\\
\enddata
\tablenotetext{a}{Hydrogen column density determined from {\tt tbvarabs} in units of $10^{22}$\,cm$^{-2}$.  The Galactic value of n$_H$ was fixed to the \citet{dic90} value with the {\tt tbabs} model.}
\tablenotetext{b}{Element abundance relative to the Wilms solar abundance from the {\tt tbvarabs} model}
\tablenotetext{c}{Ratio of mass accretion rate from the {\tt grad} model to Eddington accretion rate (see Section 4)}
\tablenotetext{d}{Total number of photon counts from pn and MOS detectors}
\end{deluxetable*}
\begin{deluxetable*}{lllllllll}
\tablecaption{Spectral Fits for RGS spectra with {\tt tbabs*tbvarabs*edge*(grad + pow)}  model\label{tbl-3}}
\tablewidth{0pt}
\tablehead{
\colhead{Source} & \colhead{n$_{H}$\tablenotemark{a}} & \colhead{O abund.\tablenotemark{b}} 
& \colhead{Fe abund.\tablenotemark{b}} & \colhead{Mass (M$_{\sun}$)} 
& \colhead{$\dot{M}/\dot{M}_{Edd}$\tablenotemark{c}} & \colhead{$\Gamma$} & \colhead{$\chi^{2}$/dof} 
& \colhead{counts\tablenotemark{d}}
}
\startdata
HolmII XMM1	& 0.08$^{+0.05} _{-0.02}$ & 0.65$^{+0.61} _{-0.64}$ & 0.0$^{+1.51}$ &147$^{+88} _{-83}$ & 0.24$^{+0.11} _{-0.12}$ & 1.58$^{+6.75} _{-4.58}$&  444.4/442 & 9521\\
\\
HolmIX XMM1	& 0.29$^{+0.12} _{-0.08}$ & 0.68$^{+0.30} _{-0.34}$ & 0.36$^{+1.36} _{-0.36}$ &774$^{+4226} _{-570}$ & 0.09$^{+0.16} _{-0.03}$ & 1.43$^{+0.45} _{-0.98}$ & 290.4/339 & 10807\\
\enddata
\tablenotetext{a}{Hydrogen column density determined from {\tt tbvarabs} in units of $10^{22}$\,cm$^{-2}$.  The Galactic value of n$_H$ was fixed to the \citet{dic90} value with the {\tt tbabs} model.}
\tablenotetext{b}{Element abundance relative to the Wilms solar abundance}
\tablenotetext{c}{Ratio of mass accretion rate from the {\tt grad} model to Eddington accretion rate (see Section 4)}
\tablenotetext{d}{Total number of photon counts from RGS1 and RGS2 detectors}
\end{deluxetable*}

For M33 X-8, the source was well-fit
with an absorbed blackbody
and power law with best-fit parameters: n$_H = 1.67^{+0.09}_{-0.08} \times 10^{21}$\,cm$^{-2}$, kT$ = 0.74^{+0.02}_{-0.02}$\,keV, $\Gamma = 2.46^{+0.06}_{-0.05}$, 
and $\chi^2 = 1579.8/1533$ dof, where n$_H$ represents the host galaxy/ULX hydrogen column (the Milky Way
contribution was fixed to the \citet{dic90} value listed in Table~\ref{tbl-5}).  
In \citetalias{win05} we noted that ULX sources well-fit with the combined blackbody
and power law model with a higher disk temperature ($\approx 1$\,keV) and a lower flux were often well-fit
by an absorbed Comptonization model (XSPEC model {\tt compst}).  We note that for M33 X-8
the inverse Compton scattering model also fits the data well 
(wabs*compst) but with a $\chi^2 = 1614.9/1536$ dof.  
Despite the larger $\chi^2$
value, the {\tt compst} model better fits the residuals from the spectra's sloping high energy tail.    
The best-fit absorbed blackbody and power law parameters for 
NGC 4559 X7 (12\,h\ 35\,m\ 51.8\,s, $27^{\circ}\ 56^{\prime}\ 4^{\prime\prime}$) were:
n$_H = 1.51^{+0.04}_{-0.01} \times 10^{21}$\,cm$^{-2}$, kT$ = 0.13^{+0.01}_{-0.02}$\,keV, $\Gamma = 2.16^{+0.10}_{-0.04}$, and $\chi^2 = 410.3/369$ dof.  For
NGC 4559 X10 (12\,h\ 35\,m\ 58.6\,s, $27^{\circ}\ 57^{\prime}\ 40.8^{\prime\prime}$), we found:
n$_H = 1.14^{+0.02}_{-0.01} \times 10^{21}$\,cm$^{-2}$, kT$ = 0.96^{+0.26}_{-0.20}$\,keV, $\Gamma = 2.13^{+0.06}_{-0.05}$, and $\chi^2 = 292.3/330$ dof.  
X10, as noted in \citet{cro04}, is also well fit by a Comptonization model.

For the sources M33 X-8 and NGC 4559 X10, we noted that their spectra were well-fit by either an
absorbed blackbody and power law or an absorbed Comptonization model.  This brings up one
issue surrounding ULX spectra.  Namely, ambiguity over their spectral form.  In this paper, we assume
that there is an analogy between ULXs and Galactic black hole binaries.  Thus, we assume that the 
accurate ULX spectrum for these bright sources is a hard component (well modeled by a power law) 
and a soft component (which we assume as a thermal contribution from an accretion disk).  The form of
the soft component, in particular, will affect the measured absorption and abundance values.  Thus,
an investigation of this component and its affect on the absorption model is important.
 
 \begin{figure}
\plotone{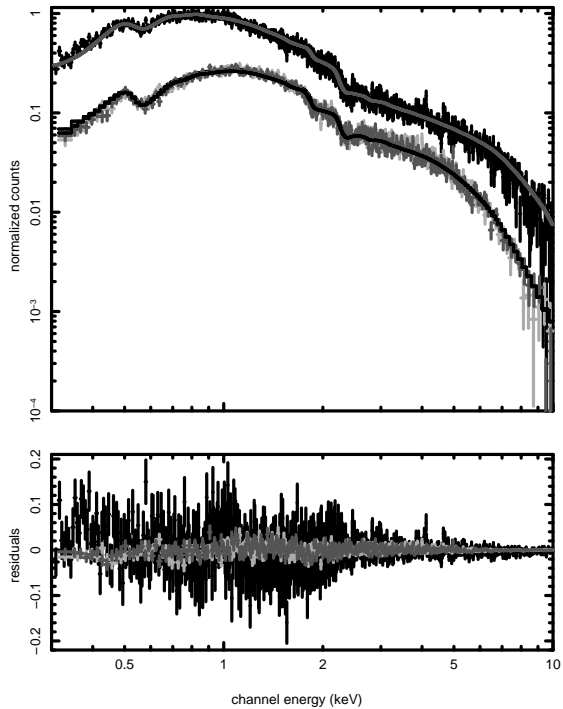}
\caption{EPIC spectrum of observation 0200980101 of Holmberg IX XMM1.  This plot
shows the spectrum fit with the {\tt tbvarabs*tbabs*edge*(grad + pow)} model and
the residuals from this fit.  The best fit
parameters are listed in Table~\ref{tbl-2}.
\label{fig-spectrum}}
\end{figure}
 
The soft component of ULXs is most often
modeled as a thermal component originating from an accretion disk surrounding a central black
hole.   There are numerous disk models applied to model this possibly thermal component.  In
\citetalias{win05}, we modeled this component as a simple blackbody (in XSPEC {\tt bbody}).  
While an accretion disk is expected to have a range of temperatures,
empirically a single blackbody is a good fit to low signal-to-noise spectra.  A simple absorbed blackbody
and power law model was used for Galactic black hole X-ray binaries in the 1980s when the
quality of data for these sources was analogous to that for ULX sources today.  As the
next step in accretion disk models, the soft component of ULX spectra is often modeled as an
optically thick, geometrically thin, multi-component blackbody disk (a multi-component disk or 
MCD model, {\tt diskbb} in XSPEC) \citep{mit84}. 

Two disk models that are used to fit the soft component of ULX spectra with more physical accuracy are
the XSPEC models {\tt diskpn} and {\tt grad}.  The {\tt diskpn} model is an extension of the MCD ({\tt diskbb})
model which uses a pseudo-Newtonian potential. 
The {\tt grad} model \citep{han89, ebi91} is a multi-component disk model that unlike the {\tt diskbb} or {\tt diskpn} model, incorporates the affects
of general relativity.  One of the advantages of the {\tt grad} model is that it fits the spectra for  mass
(M$_{grad}$) and mass
accretion rates ($\dot{M}$) given a few initial assumptions (distance to the source, disk inclination angle, and the ratio of
the color temperature to the effective temperature).  

In addition to thermal models, other models have been suggested to explain the soft component. 
Two of these models are ionized reflection and the warm absorber model.  Both of these models
have been applied to low-redshift PG Quasars, sources with blackbody temperatures of 150\,eV.
This is relevant to ULXs since many spectral fits of ULXs
require cool accretion disk temperatures of approximately 100\,eV \citep{mil03, mil04,rob05, win05}.
The reflection model suggests that the soft component results from X-ray ionized reflection.  
In this model, back-scattering and fluorescence of X-rays in the disk, 
as well as radiative recombination, cause elements 
with smaller ionization potentials (e.g. C, O, N) to become highly ionized.  
\citet{ros05} note that a relativistically
blurred X-ray ionization model folded through an {\it XMM-Newton} pn response matrix,
is well-fit by a blackbody with a temperature of 150\,eV, the same value that is seen in PG Quasars
and many ULXs.  In the warm absorber model, absorption edges
and lines from an absorbing material close to the X-ray source appear as a thermal 
component in poorer quality spectra.  With increased spectral resolution, the numerous absorption
edges and lines are distinguishable.  The warm absorber may be the result of a strong, mildly relativistic
wind from the disk as suggested by \citet{gie04}.  Indeed, the presence of a warm absorber is
well known and studied in many Seyfert galaxies (e.g. NGC 3783 and MCG--6-30-15).   

As noted above, for subsequent spectral fits we assume a thermal model for the soft component of the 
ULX spectra.  In the following section (Section 4), we will discuss the effect different thermal models have on the measured
absorption values.  In order to measure the hydrogen column density, we fit the spectra of sources listed in
Table~\ref{tbl-5} with the more sophisticated Tuebingen-Boulder ISM absorption 
model of \citet{wil00} ({\tt tbabs}, {\tt tbvarabs} in XSPEC).  This model accounts for X-ray absorption resulting 
from contributions from X-ray absorption from the gas phase of the ISM, grains, and molecules.  The model uses 
updated solar abundances and photoionization cross-sections.  For both the EPIC and RGS spectra, we accounted 
for Galactic hydrogen absorption by setting the column density of the {\tt tbabs} model equal to the Milky 
Way hydrogen column density along the line of sight to the host galaxy .  These column densities are 
quoted in Table 1 and are all less than $6 \times 10^{20}$\,cm$^{-2}$.    

Since the measured
X-ray column densities for ULXs are typically an order of magnitude higher than the \citet{dic90}
Milky Way
values (Fig. 8 of \citetalias{win05}), we are confident that the additional absorption measured
is not from the Milky Way.   To determine the host galaxy's hydrogen column density and 
the abundances of elements along the line of sight (oxygen and iron),
we used the {\tt tbvarabs}
model.  The {\tt tbvarabs} model accounts for X-ray absorption due to photo-ionization.
It includes the effects due to the H$_2$ molecule and depletion of metals in grains.
The model allows for individual fits to abundant elements (He through Ni), H$_2$, and depletion
of elements in grains.  We initially allowed the hydrogen column density to vary, fixing
all other {\tt tbvarabs} parameters to their defaults.  The Galactic column density  remained
fixed (using the {\tt tbabs} model), while the other parameters (power law and blackbody components) were allowed to vary.
This was modeled as {\tt tbabs*tbvarabs*(bbody + pow)} in XSPEC, where {\tt tbabs} was fixed to the Milky Way value
and {\tt tbvarabs} was used to fit the absorption from the ULX/host galaxy contribution.
With the best-fit hydrogen column density (from {\tt tbvarabs}), we allowed the oxygen abundance and then the iron
abundance to vary from the solar abundances.  Allowing these parameters to float provides measurements of
the  depth of the oxygen K-shell edge at 542\,eV and the iron L-shell edge at 851\,eV.

As in \citet{bau05}, we found that for our 14 sources, the oxygen absorption values from the {\tt tbvarabs} model
yielded different values for the EPIC pn and MOS spectra.
Thus, we follow the procedure of \citet{bau05} in adding an {\tt edge} model to account for the differences.
We add an extra edge component to the MOS1 and MOS2 detectors at an energy of 0.53\,keV with optical depths of
0.22 and 0.20 respectively (modeled as {\tt tbabs*tbvarabs*edge*(bbody + pow)})\footnotemark.  This is the template model used for
all spectral fits mentioned throughout the rest of the paper, where other thermal models are used
in place of the simple blackbody ({\tt bbody}) model as indicated.

An additional problem we note involves the hydrogen column densities along the line of sight.  \citet{bau05} note
that a column density (n$_H$) greater than $0.5 \times 10^{21}$\,cm$^{-2}$ is necessary to allow for good
oxygen abundance measurements. We also note that at very high n$_H$ values the oxygen edge is undetectable,
where high absorption values cause the signal-to-noise to be too low to detect the edge. 

\footnotetext{The edge parameters used by \citet{bau05} were calculated for differences in
calibration using SAS version 5.4.  In the appendix, we show that the pn values using SAS 6.5
are consistent with SAS 6.0 pn+MOS derived values.  Thus, the correction quoted is valid
for our purposes with SAS version 6.0.}   

\section{Nature of the Soft Component}

In the previous section we outlined three possible origins for the soft component in ULXs (X-ray
reflection, thermal emission from an accretion disk, and a warm absorber model).  For our 
purposes, we assume a thermal component origin.  We chose the thermal model since,
as described in Paper I, the spectra of ULXs are well described with models used to
fit Galactic black hole X-ray binaries.  To gauge the effect different disk models (i.e. {\tt bbody}, {\tt diskbb}, {\tt diskpn}, {\tt grad})
may have on the measured abundances, we discuss different disk models in section 4.1.
In section 4.3, we assess the usefulness of the model (absorbed disk and
power law)
for measuring the galactic/ULX hydrogen column density and the oxygen and iron abundances by
discussing the physical plausibility of this model.  

For sources with at least 5000 counts, the 14 sources listed in Table~\ref{tbl-5}, we fitted the {\tt grad} model 
for the soft-component of the spectra (XSPEC model {\tt tbabs*tbvarabs*edge*(grad + pow)}).
 We assumed a disk inclination angle of 60 degrees
and a ratio T$_{color}/$T$_{eff}$ = 1.7 (the default value, typical of stellar mass BH sources).  
In Table~\ref{tbl-2}, we list the best-fit parameters for EPIC spectra using this model.  Rows with no source name 
indicated represent an additional observation of the previous ULX (as indicated in Table~\ref{tbl-5}).
Table~\ref{tbl-3} provides the best-fit parameters for the RGS spectra.  The important measurements to note are
the host galaxy/ULX contribution to the hydrogen column density, oxygen abundance, and iron abundance from the {\tt tbvarabs} model.  All errors
quoted in this paper are for the 90\% confidence level for one degree of freedom ($\Delta\chi^2 = 2.76$).
We provide a representative spectral fit in Figure~\ref{fig-spectrum} for the long observation of
Holmberg IX XMM1.

\subsection{Comparison of Thermal Disk Models}
We chose the general relativistic disk model, {\tt grad}, because it is the most 
physically accurate of the various simple accretion disk models.  Also, the {\tt grad}
model requires
few initial parameters while making a direct calculation of the mass and mass
accretion rate of the black hole.  In order to compare the {\tt grad} results
with an alternate model, we fit the highest signal-to-noise
observation of Holmberg IX XMM1 with the {\tt diskpn} model (with the inner radius
of the disk set to the radius of marginal stability or 6 times the Schwarzschild radius) in
place of the {\tt grad} model.  We found that the values obtained agreed with those
of the {\tt grad} model.  The mass from the {\tt diskpn} model was slightly lower
at 358\,M$_{\sun}$ (corresponding to kT$= 0.23$\,keV) and the iron abundance was slightly 
higher at 2.33 compared to the {\tt grad} model.  Since the {\tt diskpn} and {\tt grad} models
yield similar values, either of these disk models would be sufficient for determining the
hydrogen column density and oxygen and iron abundances.

\begin{deluxetable*}{llllllll}
\tablecaption{Alternate Thermal Model Spectral Fits to ULX Sources with $> 20000$\,Counts\label{tbl-1}}
\tablewidth{0pt}
\tablehead{
\colhead{Source} & \colhead{n$_{H}$\tablenotemark{a}} & \colhead{Oxygen abundance\tablenotemark{b}} & \colhead{Iron abundance\tablenotemark{b}}
& \colhead{kT} & \colhead{$\Gamma$} & \colhead{$\chi^{2}$/dof} & \colhead{counts\tablenotemark{c}}
}
\startdata
\cutinhead{{\tt tbabs*tbvarabs*edge*(bbody + pow)}}
HolmII XMM1	& 0.12$^{+0.01} _{-0.01}$ & 0.97$^{+0.11} _{-0.12}$ &  0.0$^{+0.32}$ & 0.24$^{+0.01} _{-0.01}$ & 2.41$^{+0.03} _{-0.03}$ & 2216.5/2018 & 342874\\
\\
\nodata		& 0.14$^{+0.03} _{-0.02}$ & 1.37$^{+0.38} _{-0.39}$ &  4.38$^{+0.62} _{-1.8}$ & 0.16$^{+0.01} _{-0.01}$ & 2.31$^{+0.08} _{-0.04}$ & 952.7/973 & 43116\\
\\
HolmIX XMM1	& 0.14$^{+0.02} _{-0.01}$ & 1.51$^{+0.19} _{-0.15}$ &  3.37$^{+0.99} _{-0.80}$ & 0.19$^{+0.01} _{-0.01}$ & 1.40$^{+0.02} _{-0.01}$ & 2856.7/2752 & 148061\\
\\
\nodata		& 0.23$^{+0.10} _{-0.04}$ & 1.37$^{+0.43} _{-0.44}$ &  4.75$^{+0.25} _{-1.94}$ &  0.17$^{+0.02} _{-0.03}$ & 1.71$^{+0.14} _{-0.06}$ &559.5/880 & 28108\\
\\
M33 X-8& 0.22$^{+0.01} _{-0.14}$ & 1.24$^{+0.10} _{-0.10}$ &  2.38$^{+0.31} _{-0.60}$ & 0.76$^{+0.02} _{-0.02}$ & 2.51$^{+0.07} _{-0.06}$ & 1528.8/1533 & 123903\\
\\
M81 XMM1 	& 0.39$^{+0.03} _{-0.02}$ & 1.13$^{+0.15} _{-0.07}$ &  1.78$^{+0.65} _{-0.40}$ &  0.90$^{+0.03} _{-0.02}$ & 2.70$^{+0.05} _{-0.05}$ &1240.2/1241 & 69776\\
\\
\nodata		& 0.26$^{+0.09} _{-0.04}$ & 1.46$^{+0.26} _{-0.49}$ &  1.61$^{+2.02} _{-1.61}$ &  0.99$^{+0.38} _{-0.28}$ & 1.76$^{+0.13} _{-0.09}$ & 627.4/668 & 31731\\
\\
NGC253 XMM2	& 0.22$^{+0.05} _{-0.05}$ & 1.21$^{+0.40} _{-0.37}$ &  1.53$^{+1.67} _{-1.53}$ & 0.73$^{+0.14} _{-0.11}$ & 2.16$^{+0.08} _{-0.23}$ & 460.8/496 & 20651\\
\\
\cutinhead{{\tt tbabs*tbvarabs*edge*(diskbb + pow)}}
HolmII XMM1	& 0.11$^{+0.001} _{-0.01}$ & 0.65$^{+0.09} _{-0.05}$ &  0.0$^{+0.06}$ &  0.34$^{+0.02} _{-0.02}$ & 2.38$^{+0.05} _{-0.04}$ & 2241.8/2018 & 342874\\
\\
\nodata		& 0.15$^{+0.03} _{-0.02}$ & 1.05$^{+0.34} _{-0.38}$ &  2.67$^{+1.65} _{-1.71}$ & 0.21$^{+0.04} _{-0.01}$ & 2.27$^{+0.06} _{-0.07}$ &  954.2/973 & 43116\\
\\
HolmIX XMM1	& 0.19$^{+0.02} _{-0.02}$ & 1.38$^{+0.11} _{-0.16}$ &  2.39$^{+0.58} _{-0.84}$ &0.24$^{+0.02} _{-0.01}$ & 1.39$^{+0.02} _{-0.03}$ & 2878.9/2752 & 148061\\
\\
\nodata		& 0.30$^{+0.11} _{-0.07}$ & 1.36$^{+0.32} _{-0.38}$ &  4.03$^{+0.97} _{-1.65}$ & 0.20$^{+0.04} _{-0.03}$ & 1.72$^{+0.10} _{-0.07}$ &  559.8/880 & 28108\\
\\
M33 X-8& 0.19$^{+0.04} _{-0.03}$ & 1.04$^{+0.15} _{-0.16}$ &  1.75$^{+0.68} _{-0.71}$ & 1.17$^{+0.05} _{-0.05}$ & 2.47$^{+0.21} _{-0.15}$ & 1530.4/1533 & 123903\\
\\
M81 XMM1 	& 0.63$^{+0.08} _{-0.15}$ & 1.19$^{+0.09} _{-0.15}$ &  2.07$^{+0.40} _{-0.69}$ & 1.38$^{+0.03} _{-0.08}$ & 4.08$^{+0.75} _{-0.59}$ &  1247.7/1241 & 69776\\
\\
\nodata		& 0.27$^{+0.18} _{-0.08}$ & 1.42$^{+0.57} _{-0.49}$ &  1.63$^{+1.69} _{-1.63}$ & 2.06$^{+0.55} _{-0.97}$ & 1.98$^{+1.40} _{-0.58}$ &  626.9/668 & 31731\\
\\
NGC253 XMM2	& 0.21$^{+0.10} _{-0.07}$ & 1.16$^{+0.53} _{-0.37}$ &  1.64$^{+1.64} _{-1.64}$ & 1.27$^{+0.22} _{-0.27}$ & 2.27$^{+0.83} _{-0.62}$ & 461.3/496 & 20651\\
\enddata
\tablenotetext{a}{Hydrogen column density determined from {\tt tbvarabs} in units of $10^{22}$\,cm$^{-2}$.  The Galactic value of n$_H$ was fixed to the \citet{dic90} value with the {\tt tbabs} model.}
\tablenotetext{b}{Element abundance relative to the Wilms solar abundance from the {\tt tbvarabs} model}
\tablenotetext{c}{Total number of photon counts from pn and MOS detectors}

\end{deluxetable*}

In order to test the effect the thermal model has on the {\tt tbvarabs} measured parameters
(i.e. n$_H$ and abundances), we wanted to compare the results from the more physical
disk model ({\tt grad}) with less physical models ({\tt bbody} and {\tt diskbb}).  In order to 
compare the results from these three models, we chose to examine the best-fit parameters
for spectra with the highest signal-to-noise (using sources with at least 20000\,counts).
Thus, in Table~\ref{tbl-1} we list the best-fit parameters obtained
using an absorbed blackbody and power law ({\tt tbabs*tbvarabs*edge*(bbody + pow)})
 and an absorbed {\tt diskbb} and power law model ({\tt tbabs*tbvarabs*edge*(diskbb + pow)})
 for sources with at least 20000 counts.  Subsequently in this paper, when we refer to the
 {\tt bbody}, {\tt diskbb}, or {\tt grad} models, we are referring to the model fits used in
 Tables~\ref{tbl-2}-~\ref{tbl-1}. 

 The curvature, or shape of the spectra at low energies, for the 
 {\tt bbody}, {\tt diskbb}, and {\tt grad} models is different, presumably affecting
the column density and absorption values.  From a comparison of these models,
comparing the mean values from all of the observations with $> 20000$\,counts 
listed in Tables~\ref{tbl-2} and~\ref{tbl-1},
we find that the column densities obtained with the {\tt diskbb} model are nearly 
identical to those with the {\tt grad} model (the exception is the source NGC 253 XMM2).
The MCD ({\tt diskbb}) model hydrogen column densities are approximately 6\% higher than those of
the blackbody model (excluding the first M81 XMM1 observation).
The average power law indices for the three models agree within a factor of 5\%, with the
{\tt bbody} model having the lowest and the {\tt grad} having the highest
average power law index (excluding the first M81 XMM1 observation).  Thus, the difference
in both power law index and hydrogen column density is negligible on average, with the
exceptions of NGC 253 XMM2 and M81 XMM1.  

For a comparison of the derived abundances, the oxygen abundance is roughly the same between
the {\tt grad} and {\tt bbody} model.  These values are
approximately 10\% higher than those derived from the MCD ({\tt diskbb}) model.  The iron abundances, however,
vary more from model to model.  The blackbody model average iron
abundance is $\approx 13$\% higher than that of the {\tt grad} model and
$\approx 18$\% higher than that of the {\tt diskbb} model.  Thus, there are no significant 
changes to the model parameters in changing the base accretion disk model.  These results
are verified by simulations described in the appendix.
\begin{figure}
\plotone{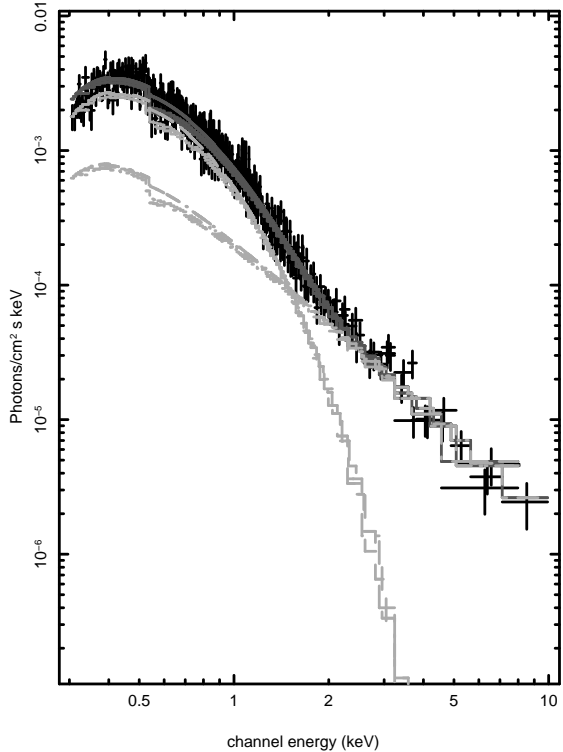}
\caption{ Plot of the EPIC unfolded spectrum for the NGC 5408 XMM1 ULX with
the high mass solution (see Table 2).  With the spectrum, the absorbed {\tt grad} and 
absorbed power law model components are plotted with light gray lines while the
combined model is plotted in dark gray.  The thermal component clearly dominates
the low energy spectrum with this model.  
\label{fig-high}}
\end{figure}

\begin{figure}
\plotone{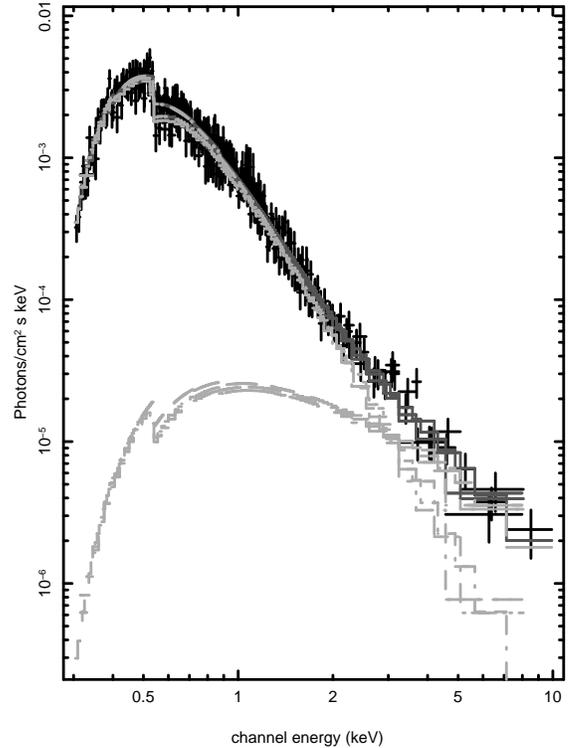}
\caption{ Plot of the EPIC unfolded spectrum for the NGC 5408 XMM1 ULX with
the low mass solution (see Table 5).  With the spectrum, the absorbed {\tt grad} and 
absorbed power law model components are plotted with light gray lines while the
combined model is plotted in dark gray.  Notice that the low energy spectrum is dominated by
the power law component with a weak contribution from the disk model ({\tt grad}).
\label{fig-low}}
\end{figure}

\subsection{Degenerate Model}

In modeling our sources with the {\tt tbvarabs*tbabs*edge*(grad+pow)} model, we found
that some of the ULX sources were well-fit by a model where the power law component
dominates the low energy spectrum, also seen in \citet{sto06}.   
This type of spectral fit typically yields a steeper
power law index ($\Gamma > 3.0$) and a low mass (M$< 10$\,M$_{\sun}$).  For
two sources, M33 X-8 and M81 XMM1, this model fit was a much better fit than a higher mass
model with $\Delta\chi^2$ of 476 and 190 respectively.  However, for some sources,
there was a degeneracy between the two models (high mass and low mass/steep power law).
We illustrate this degeneracy with observation 0112290601 of the source NGC 5408 XMM1, 
showing the high mass model
and spectral fit in Figure~\ref{fig-high} and the low mass model and spectral fit in Figure~\ref{fig-low}.  For
spectra exhibiting this degenerate solution, we include the low-mass/steep power law fits
in Table~\ref{tbl-deg}.  These fits all exhibit, in addition to low masses, solutions with 
$\dot{M} >> \dot{M}_{Edd}$.

\begin{deluxetable*}{lllllllll}
\tablecaption{Spectral Fits for EPIC spectra with {\tt tbabs*tbvarabs*edge*(grad + pow)} model
for Sources with Degenerate Solution (Low Mass/Steep Power Law)\label{tbl-deg}}
\tablewidth{0pt}
\tablehead{
\colhead{Source} & \colhead{n$_{H}$\tablenotemark{a}} & \colhead{O abund.\tablenotemark{b}} 
& \colhead{Fe abund.\tablenotemark{b}} & \colhead{Mass (M$_{\sun}$)} 
& \colhead{$\dot{M}/\dot{M}_{Edd}$\tablenotemark{c}} & \colhead{$\Gamma$} & \colhead{$\chi^{2}$/dof} 
& \colhead{counts\tablenotemark{d}}
}
\startdata
HolmII XMM1& 0.20$^{+0.04}_{-0.04}$ & 0.83$^{+0.17}_{-0.27}$ & 1.11$^{+0.83}_{-1.11}$ & 2.24$^{+1.52}_{-1.24}$ & 7.79$^{+1.52}_{-1.95}$ &3.10$^{+0.31}_{-0.30}$ & 0.99 & 43116\\
\\
HolmIX XMM1&  0.39$^{+0.13}_{-0.12}$ & 1.05$^{+0.21}_{-0.33}$ & 2.08$^{+0.95}_{-1.45}$ & 3.91$^{+1.47}_{-1.59}$ & 22.0$^{+1.79}_{-2.86}$ & 3.51$^{+0.79}_{-0.87}$ & 0.64 & 28108\\
\\
NGC253 XMM2 & 0.18$^{+0.17}_{-0.08}$ & 1.10$^{+0.42}_{-0.55}$ & 1.32$^{+1.90}_{-1.32}$ & 6.82$^{+4.19}_{-2.22}$ & 1.27$^{+0.71}_{-0.40}$ & 2.15$^{+0.84}_{-0.58}$ & 0.93 & 20651\\
\\
NGC5204 XMM1& 0.18$^{+0.05}_{-0.04}$ & 1.00$^{+0.30}_{-0.40}$ & 0.0$^{+0.59}_{-0.0}$ & 3.21$^{+0.98}_{-0.95}$ & 6.86$^{+0.67}_{-0.09}$ & 3.28$^{+0.39}_{-0.38}$ & 0.95 & 16717\\
\\
\nodata & 0.20$^{+0.06}_{-0.05}$ & 0.63$^{+0.35}_{-0.48}$ & 0.0$^{+0.62}_{-0.0}$ & 2.95$^{+2.13}_{-1.84}$ & 7.37$^{+1.35}_{-2.08}$ & 3.16$^{+0.50}_{-0.44}$ & 0.94 & 13864\\
\\
NGC300 XMM1 & 0.21$^{+0.05}_{-0.03}$ & 1.59$^{+0.22}_{-0.26}$ & 0.0$^{+0.53}_{-0.0}$ & 1.69$^{+0.85}_{-0.32}$ & 0.73$^{+0.18}_{-0.13}$ & 3.89$^{+0.21}_{-0.32}$ & 1.00 & 11479\\
\\
N4559 X-7 & 0.20$^{+0.06}_{-0.05}$ & 0.34$^{+0.38}_{-0.34}$ & 0.0$^{+0.48}_{-0.0}$ & 5.79$^{+2.82}_{-2.75}$ & 6.63$^{+1.44}_{-2.42}$ & 3.13$^{+0.55}_{-0.50}$ & 0.83 & 12506\\
\\
NGC5408 X-1 & 0.18$^{+0.05}_{-0.04}$ & 0.87$^{+0.28}_{-0.42}$ & 0.45$^{+1.75}_{-0.45}$ & 2.29$^{+1.57}_{-1.28}$ & 4.87$^{+1.23}_{-1.02}$ & 4.09$^{+0.28}_{-0.34}$ & 0.89 & 10045\\
\enddata
\tablenotetext{a}{Hydrogen column density determined from {\tt tbvarabs} in units of $10^{22}$\,cm$^{-2}$.  The Galactic value of n$_H$ was fixed to the \citet{dic90} value with the {\tt tbabs} model.}
\tablenotetext{b}{Element abundance relative to the Wilms solar abundance from the {\tt tbvarabs} model}
\tablenotetext{c}{Ratio of mass accretion rate from the {\tt grad} model to Eddington accretion rate (see Section 4)}
\tablenotetext{d}{Total number of photon counts from pn and MOS detectors}
\end{deluxetable*}

In \citet{sto06}, the authors discussed the same issues in fitting the XMM spectra of 13 ULXs.  
They found that two sources, M33 X-8 and NGC 2403 X-1, were best fit by the
model with a power law fit to the low energy portion and thermal model at higher energy.
They also indicated six sources where an ambiguity existed between the two models.  To
understand the spectra of the sources where both models provided good fits to the data, we
further investigated the spectra of sources with multiple observations. 

For Holmberg II X-1 and Holmberg IX X-1 our spectral fits include multiple
observations (a shorter and a 100\,kilo\,second observation).  For each of these sources, we found
that the shorter observation could be fit with either a high mass or low mass solution.  When
we fit the 100\,kilo\,second observation, however, the high mass model was a much better fit.
The $\Delta\chi^2$ values between the high mass and low mass solutions for the kilo\,second 
observations were 135 and 88, respectively.  To test this further, we also fit a 100\,kilo\,second
observation of NGC 5408 XMM1 with both of these models.  This observation (0302900101)
is a proprietary observation whose spectral and temporal analysis will appear in Strohmayer
{\it et al.} (in prep).  We processed the pn data with SAS 6.5, following the same procedure
as noted in the Data Reduction section.  Fitting the spectrum with both models (high mass
and low mass solution) we found a $\Delta\chi^2$ value of 170, favoring the high mass model.

Thus, we find that for sources that are well fit by either model (showing a degenerate solution
of either high mass or low mass/steep power law) the high mass solution is the best fit when
a higher count spectrum is obtained (as for Holmberg IX XMM1, Holmberg II XMM1, and
NGC 5408 X-1).  Though we list the alternate model column density and absorption values
in Table~\ref{tbl-deg}, we use the parameters listed in Table~\ref{tbl-2} throughout the paper
(the high mass solutions). 

As noted, M33 X-8 and M81 XMM1, sources with very high number of counts, were not well-fit
with a standard disk at low energy, power law at high energy model.  They were best fit
with the steep low energy power law and hot disk model shown in Figure~\ref{fig-low}.  
Along with these sources, NGC 4559 X-10
and M83 XMM1 were also well-fit by this model.  We will further discuss these sources in 
the following subsection.

\begin{figure}
\plotone{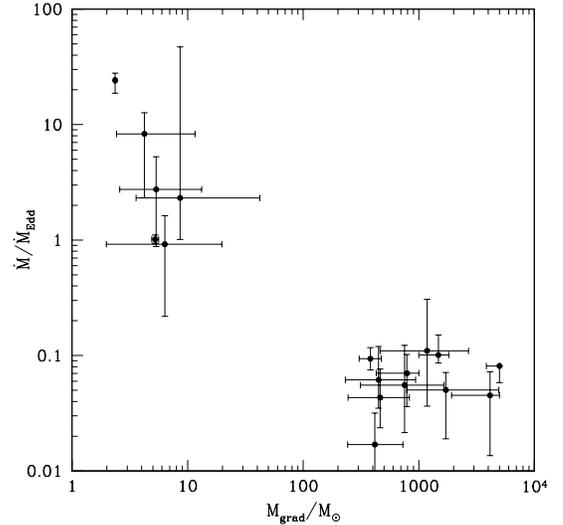}
\caption{Black hole mass versus mass accretion rate (per Eddington accretion rate) obtained
from the general relativistic disk model ({\tt grad}).  The sources with model parameters
indicating high masses correspond to accretion rates that are below 40\% of the Eddington 
accretion rate for
the given mass.  In Paper 1, we noted a temperature gap in the range of 0.26\,keV to 0.50\,keV
from spectral fits utilizing an absorbed blackbody and power law model.  In this figure, we
see that there is a gap in mass from the parameters of the absorbed general relativistic disk
and power law model.
The values shown are results for the EPIC spectra only, as recorded
in Table 2 (for all sources in Table 2).  Therefore, there are multiple points for the sources
with multiple observations.  We plotted these sources (Holmberg II XMM1,
Holmberg IX XMM1, M81 XMM1, and NGC 5204 XMM1) twice to confirm that their masses,
as determined by the {\tt grad} model, do not vary.  
\label{fig8}}
\end{figure}

\subsection{Physical Plausibility of the Accretion Disk models}
In Paper 1, we had found that ULX spectra are consistent with the high/soft and low/hard
states of Galactic black holes.  We had classified the sources in this study as consistent
with the high/soft state.  The high/soft state, as stated earlier, is characterized by emission
from an accretion disk and a Comptonized power law tail.  In order to investigate whether
the mass and accretion rate results from the {\tt grad} model make physical sense in
terms of the X-ray binary model, we present a comparison of the derived accretion
rate in Eddington accretion rate versus the mass in Figure~\ref{fig8}.  The parameter $\dot{M}_{Edd}$ was
computed as $\dot{M}_{Edd} = \frac{1.3\times10^{38} M}{\eta c^2}$ where $\eta$, the efficiency
factor,
was set to 0.06 and M is the value M$_{grad}$.  We find that for the sources with
M$_{grad} > 100$\,M$_{\sun}$, the accretion rate is computed to be below 40\% of
the Eddington rate.  This is assuming, as the {\tt grad} model assumes, a Schwarzschild black hole.  
Noting that $\dot{M}  / \dot{M}_{Edd}$ is equivalent to L\,$/$\,L$_{Edd}$ for most disk solutions, the 
L\,$/$\,L$_{Edd}$ values
for the sources with M$_{grad} > 100$\,M$_{\sun}$ are consistent with those of Galactic
black hole X-ray binaries in the high state (L\,$/$\,L$_{Edd} \approx 0.05 - 1.0$) \citep{mcc04}.
Thus, the sources with M$_{grad} > 100$\,M$_{\sun}$ do have accretion rates that are
predicted from scaling up (in mass) observed high state Galactic black holes.  In addition,
the spectral fit parameters for NGC 4631 XMM1, with an estimated mass of 5.5\,M$_{\sun}$ and an accretion rate
L\,$/$\,L$_{Edd}$ of 0.84, are also consistent with the standard high state Galactic black hole
model.  This source is likely a normal stellar mass black hole X-ray binary in an external
galaxy.

The remaining sources with M$_{grad} << 100$\,M$_{\sun}$ (M33 X-8, M81 XMM1, NGC 4559 X-10,
and M83 XMM1) yielded L\,$/$\,L$_{Edd}$ ratios in the range of 1 - 3.  These sources are
also those described in the previous section where the power law component fits the low energy
spectrum.  They are also well fit by a Comptonization model and correspond to a sub-class of
 high luminosity ULXs described in \citetalias{win05}.  Due to the luminosity
and modeled disk temperature (kT $\approx 1$\,keV), we suggested that these sources 
were very high state stellar
mass black hole systems.  In order to be consistent with the black
hole accretion model assumed in this study, the spectra of these sources should 
be the result of Comptonization
from a thermal disk spectrum.  

To test this further, we fit the spectrum of the highest count source
of this type (M33 X-8) with an absorbed thermal disk and Comptonization model 
({\tt tbabs}*{\tt tbabs}*({\tt diskpn} + {\tt comptt})).  The {\tt comptt} model has the parameters: a seed
temperature (keV), a plasma temperature (keV), and optical depth of the medium.  We used
the {\tt diskpn} model in place of the {\tt grad} model since the former provides a disk temperature
to which the {\tt comptt} seed temperature can be fixed.  This provides a physical model, where
the thermal disk supplies the energy for the Compton tail.  M33 X-8 was well fit by this model
with a $\Delta\chi^2$/dof = 1659.9/1534 (1.08).  Thus, these sources are still consistent with
the black hole accretion model.   Replacing the {\tt tbabs} model used for the galactic column
density with the {\tt tbvarabs}, we measured the column density, oxygen abundance, and iron
abundance.  With this thermal disk and Comptonization model we obtained
n$_H = 1.5^{+0.04}_{-0.08}\times10^{21}$\,cm$^{-2}$, O/H $= 1.26^{+0.08}_{-0.15}$, 
and Fe/H $= 2.83^{+0.71}_{-0.76}$, with a $\Delta\chi^2$/dof = 1577.3/1534 (1.03).  Within the
error bars, these results are consistent with those seen in Table~\ref{tbl-2}.

\section{Properties of the ISM in ULX Host Galaxies}

The major question to be examined in using ULXs as probes of the ISM is whether the hydrogen column density and
element abundances are primarily from the host galaxy or intrinsic to the local environment of the ULX.  
Before we can answer this question, it is important to understand the intrinsic spectrum of the source.
In the previous section we discussed the nature of the soft component in light of the high signal-to-noise spectra of the 14 ULX sources we examined.  
We found that if the spectrum is due to thermal emission from a disk, modeling the spectrum with
a variety of disk models ({\tt grad}, {\tt diskbb}, {\tt diskpn}, {\tt bbody}) does not significantly change 
the measured oxygen abundance or hydrogen column density.

Assuming the reliability of the hydrogen column density and abundance measurements, based 
on their model independent values, we investigate the source of the absorption in ULX spectra.
 In order to determine whether the model n$_H$ values suggest the necessity of extra local absorption, we compare the model values
with column densities obtained from \ion{H}{1} studies.  We investigate this in section 5.1.  
The oxygen abundances (as an indication of metallicity), which we examine in
section 5.2, can provide further clues of whether the absorption we see in the X-ray spectrum is
intrinsic to the source.  
We also examine possible connections between the host galaxy's star formation rate 
and elemental abundances.  
\begin{figure}
\plotone{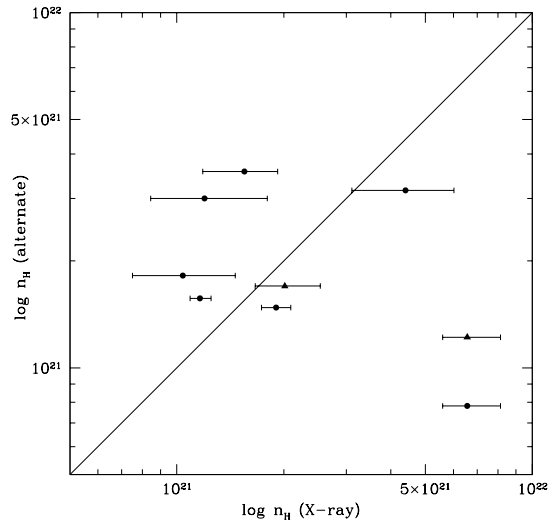}
\caption{Hydrogen column density from E$_{B-V}$ (triangle) or \ion{H}{1} studies (circle) vs. the hydrogen column density 
obtained from the {\tt tbvarabs} model.  \ion{H}{1} column densities were obtained for Holmberg II, NGC 5204, and NGC 4559 through
the WHISP survey (Swaters {\it et al.} 2002).  Additional \ion{H}{1} column densities were obtained for NGC 247, Holmberg IX, and 
M81 through VLA data (Braun 1995).  The X-ray spectral fit
columns are not biased towards significantly higher values than the alternate method column
densities.  This implies that most of the matter in the line of sight is from \ion{H}{1}.  This is not true, however, for the source M81 XMM1
(represented by the two outlying points in the lower left portion of the graph) where the 
X-ray column density is much greater than those from the optical and radio.   
\label{fig7}}
\end{figure}
\subsection{Column Densities}
To determine whether the ULX X-ray hydrogen column densities represent largely galactic 
column densities or column densities local to the ULX, we compared the 
X-ray values to those obtained from optical and radio studies.   For a comparison to hydrogen column densities from optical studies, we used interstellar reddening values.
In a study of dust scattering X-ray halos surrounding point sources and supernova remnants, \citet{pre95}
derived a relationship between hydrogen column density and interstellar reddening, using
X-ray data.  They found that
n$_H = 5.3 \times 10^{21} cm^{-2} E_{B-V}$.  They also found that these X-ray derived 
column densities are not affected
by the intrinsic absorption of the X-ray source.  Thus,  the optical reddening, $E_{B-V}$,
becomes a useful tool in checking our own X-ray
derived column densities.  Through a literature search, we found E$_{B-V}$ values for the sources M33 X-8
(0.22; \citet{lon02}) and M81 XMM1 (0.23; \citet{kon00}).  The corresponding $E_{B-V}$-derived
n$_{H}$ values are plotted 
in Figure~\ref{fig7} as triangles.

\begin{deluxetable}{lll}
\tablecaption{Column Densities from Optical and \ion{H}{1} Studies\label{tbl-6}}
\tablewidth{0pt}
\tablehead{
\colhead{Source} & \colhead{n$_{H}$\tablenotemark{a}} & \colhead{Method\tablenotemark{b}} 
}
\startdata
NGC 247 	XMM1& 0.316	& H I \\
M33 X-8		& 0.170 	& E$_{B-V}$ \\
HolmII XMM1	& 0.157	& H I \\
M81 XMM1	& 0.122 	& E$_{B-V}$ \\
M81	XMM1	& 0.078 	& H I \\
HolmIX XMM1	& 0.148	& H I \\
NGC 4559 X7 	& 0.357	&  H I \\
NGC 4559 X10	& 0.300 	&  H I \\
NGC 5204 XMM1& 0.182	& H I \\
\enddata
\tablenotetext{a}{Hydrogen column density (galactic, not Milky Way) determined by the method indicated in units of $10^{22}$\,cm$^{-2}$.}
\tablenotetext{b}{Column densities computed from reddening values (E$_{B-V}$) or from
radio H I measurements, see section 5.1 for details.}
\end{deluxetable}

 For a comparison of X-ray derived hydrogen column densities with radio values, 
 we obtained \ion{H}{1} column densities for four objects 
(Holmberg II XMM1, NGC 4559 X7, NGC 4559 X10, and NGC 5204 XMM1) from the WHISP
catalog \citep{swa02}.  These are radio \ion{H}{1} column densities within the 
host galaxy (galactic), not the Galactic/Milky Way columns.
Exact values of the \ion{H}{1} column densities were computed and given to
us by Rob Swaters.  Additionally, we include \ion{H}{1} column densities of NGC 247 XMM1, M81 XMM1, and
Holmberg IX XMM1 from \citet{bra95}.  We obtained the FITS files of \ion{H}{1} column density maps from
this paper (available on NED), where the pixel value corresponds to the galactic 
column density in units of
$10^{18}$\,cm$^{-2}$. The \ion{H}{1} n$_H$ 
values from both studies are plotted in Figure~\ref{fig7} as circles.  The column densities, from the
radio and reddening studies, are listed in Table~\ref{tbl-6}.   

We find that the host galaxy column densities from alternate
methods (optical or radio studies) are not significantly different from the X-ray column densities.  
Particularly, the X-ray values are not skewed towards substantially higher values than the
optical/radio values.
Thus, the X-ray columns are likely the galactic values without any additional local absorption.  The exception, 
however, is M81 XMM1 (represented by 2 points in Figure~\ref{fig7}, one for each of the methods).  
The E$_{B-V}$ value ($1.22\times10^{21}$\,cm$^{-2}$) and the \ion{H}{1} value 
($0.78\times10^{21}$\,cm$^{-2}$) are significantly lower than the X-ray column density.  This may indicate the presence of extra absorption around this source.   

The result that the X-ray hydrogen column densities are in good agreement with those from \ion{H}{1} studies and interstellar extinction values is interesting considering that the X-ray measured column densities are along a direct line of
sight to the ULXs while the \ion{H}{1} measurements are an average over a larger beam area.  The agreement between
the two measurements implies that the ULX sources, with the exception
of M81 XMM1, lie within roughly normal areas of their host
galaxies (i.e. not in regions of higher column density such as a molecular cloud).

\subsection{Elemental Abundances}

\subsubsection{Test for Oxygen Ionization Level}

Before discussing implications of the determined oxygen and iron abundances, 
we relate a test performed to determine whether we could distinguish between
different ionization levels of oxygen.  To do this, we used the absorption edge model, {\tt edge},
in XSPEC (using the full model: {\tt tbabs*tbvarabs*edge*edge*(grad + pow)}).  
We first checked to see that the abundance values obtained with the edge model
matched the values from the {\tt tbvarabs} model.  We fixed the oxygen abundance in the
{\tt tbvarabs} model at zero and added the edge model, allowing the threshold energy 
and absorption depth ($\tau$) to float as free parameters (with an initial energy set to 0.543\,keV).
We fit this model to the longest observations
of Holmberg II XMM1 and Holmberg IX XMM1 in addition to a source with a lower number of counts,
NGC 5408 XMM1 (0112290601).

  To find the oxygen column density (n$_O$),
we used the relationship that $\tau = \sigma \times n_O$, where $\sigma$ is the cross section for
photoabsorption.  We used the cross section values for neutral oxygen published in \citet{rei79} 
as an estimate.   This choice is supported by the results of \citet{jue04}, who measured the ratio of
oxygen ionization states in the ISM as \ion{O}{2}/\ion{O}{1}$\approx 0.1$.
In Table~\ref{tbl-4}, the hydrogen column density, threshold energy, and
optical depth are listed for the sources fit with this model.  From \citet{rei79} we used
the cross section values of  $\sigma = 5.158 \times 10^{-19}$\,cm$^2$ (for $E = 0.540$\,keV) 
and $\sigma = 4.789 \times 10^{-19}$\,cm$^2$ (for $E = 0.570$\,keV).    
As seen in Table~\ref{tbl-4}, the hydrogen column densities and the oxygen
abundances obtained from this model are close to those from the {\tt tbvarabs} model.  The [O/H]
values, where [O/H] = 12 + log(O/H) and O and H represent oxygen column density and hydrogen
column density respectively, between the two models vary by less than 1\%.  

To test whether
the threshold energy from the edge model is affected by the ionization level of oxygen, we simulated
spectra of an absorbed power law model with an oxygen edge.  Tim Kallmann (P.C.) provided
us with an oxygen edge model incorporating the cross sections of \citet{gar05}.  The model
allows for a variation of the ratio of \ion{O}{2}/\ion{O}{1}.  Using the response and ancillary response matrices
from the long Holmberg II ULX observation, we simulated spectra with the XSPEC {\tt fakeit}
command for a $\Gamma = 2.35$ power law.  We simulated spectra for \ion{O}{2}/\ion{O}{1} ratios of
0.0, 0.2, 0.4, 0.6, 0.8, and 1.0.  The simulated spectra were binned with a minimum of 20 counts/bin.
The absorption edge component of the spectrum was then fit with the {\tt edge} model.  
The fits to the threshold energies for the simulated spectra yielded values ranging 
from 0.53-0.59\,keV,  with no
preference of lower \ion{O}{2}/\ion{O}{1} ratios corresponding to lower threshold energies.  Since
the \ion{O}{1} absorption edge occurs at E$= 0.543$\,keV and the \ion{O}{2} absorption edge occurs at
E$=0.57$\,keV, we could not distinguish between these ionization states of
oxygen.  Our simulations show that the oxygen {\tt edge} measurements will be 
sensitive to \ion{O}{1} and
\ion{O}{2} but not to high ionization states (for instance \ion{O}{8} which has
an edge energy of 0.87\,keV).
\begin{figure}
\plotone{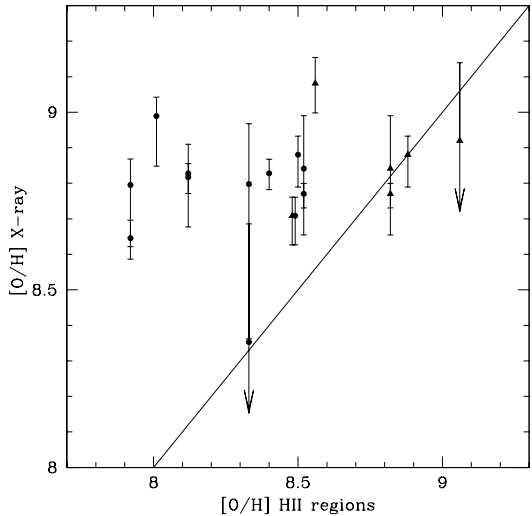}
\caption{Oxygen abundances from studies of \ion{H}{2} regions vs. oxygen abundances from the {\tt tbvarabs} model.  The [O/H]
ratios are given for Holmberg II XMM1, NGC 5408 XMM1, Holmberg IX XMM1, NGC 4559 XMM1, NGC 4559 XMM2, NGC 1313 XMM3, M33 X-8, NGC 300 XMM1, 
NGC 253 XMM2, 
and M81 XMM1 using the P-calibration method (circle).  Also, [O/H] ratios are included for M33 X-8, NGC 300 XMM1, M81 XMM1, NGC 253 XMM2,
and M83 XMM1 using the R$_{23}$ calibration method (triangle).  Arrows indicate that the lower limit
for the [O/H] X-ray parameter is below the plotted graph region.
Our values are high compared to the P-method but in good agreement
with the R$_{23}$ method.  The sources with the largest difference between [O/H] values are those located in irregular galaxies.
\label{fig2}}
\end{figure}

\subsubsection{X-ray/optical [O/H] Comparison}

As noted above, we tested the oxygen abundances obtained with the absorption model {\tt tbvarabs} against the abundances obtained from adding a photo-electric absorption edge model, for three of the ULX sources.  We found good agreement ($<  1$\% difference in [O/H] values) between both models for the X-ray spectra.  However, we found that it is not possible to distinguish between different low
ionization states of oxygen using the edge model.

\begin{deluxetable*}{llllllll}
\tablecaption{Spectral Fits for Oxygen Edge with {\tt tbabs*tbvarabs*edge*edge*(grad + pow)} model\label{tbl-4}}
\tablewidth{0pt}
\tablehead{
\colhead{Source} & \colhead{n$_{H}$\tablenotemark{a}} & \colhead{E\tablenotemark{b}} 
& \colhead{$\tau$\tablenotemark{c}} & \colhead{n$_{O}$\tablenotemark{d}} 
& \colhead{[O/H]\tablenotemark{e}}  & \colhead{$\chi^{2}$/dof} 
& \colhead{counts\tablenotemark{f}}
}
\startdata
HolmII XMM1	& 0.16$^{+0.01} _{-0.01}$ & 0.566$^{+0.01} _{-0.01}$ & 0.35$^{+0.02}_{-0.01}$ & 7.2  & 8.65/8.64 &  2543/2017 & 342874\\
\\
HolmIX XMM1	& 0.19$^{+0.02} _{-0.02}$ & 0.543$^{+0.01} _{-0.01}$ & 0.62$^{+0.11} _{-0.12}$ & 12.0 & 8.81/8.82  & 2887/2751 & 148061\\
\\
NGC 5408 XMM1 & 0.17$^{+0.06} _{-0.03}$ & 0.538$^{+0.04} _{-0.02}$ & 0.37$^{+0.29} _{-0.21}$ & 7.1 & 8.61/8.63 & 298/334 & 10045\\
\enddata
\tablenotetext{a}{Hydrogen column density determined from {\tt tbvarabs} in units of $10^{22}$\,cm$^{-2}$.  The Galactic value of n$_H$ was fixed to the \citet{dic90} value with the {\tt tbabs} model.}
\tablenotetext{b}{Threshold Energy obtained from the {\it edge} model in keV.  Note that one {\tt edge}
model was used to correct for the difference in the oxygen edge between the pn, MOS1, and MOS2
CCDs.  The other {\tt edge} model was used to measure the oxygen abundance from the
542\,eV K-shell edge.}
\tablenotetext{c}{Absorption depth obtained from the {\it edge} model}
\tablenotetext{d}{Column density of oxygen estimated from the {\it edge} model in units of $10^{17}$\,cm$^{-2}$}
\tablenotetext{e}{Abundance of oxygen relative to hydrogen from the {\it edge} model versus the value quoted in Table~\ref{tbl-2}, [O/H] = 12 + log(O/H) where O is oxygen abundance and H is hydrogen abundance }
\tablenotetext{f}{Total number of photon counts from the pn and MOS detectors}
\end{deluxetable*}
We now discuss comparisons of our X-ray oxygen absorption values with measurements in different wavelengths, based on a literature search for [O/H] ratios. 
In Figure~\ref{fig2} we compare our [O/H] ratios with those of a study conducted by \citet{pil04} (circles).  They provide a compilation of [O/H] ratios determined through spectrophotometric studies of
\ion{H}{2} regions.  Their [O/H] values are based on the radial distribution of oxygen abundance using the P-method.
They determined [O/H] values for spiral galaxies where published spectra were available for at least 4 \ion{H}{2} regions.
In addition, they reference [O/H] values for irregulars obtained through alternate methods.
Our [O/H] values are determined from the oxygen abundances listed in Table~\ref{tbl-2}.  
Thus, [O/H]$= 12 + \log(O\times0.00049)$. O is the oxygen abundance obtained from the model, which is 
multiplied by the Wilms relative abundance of $4.9\times10^{-4}$ of oxygen to hydrogen.  We were able
to compare [O/H] values for the sources located in M33, NGC 253, NGC 300, M81, Holmberg II, NGC 4559, and
NGC 5408.  We include the [O/H] value computed for the Holmberg IX ULX by \citet{mil95} of 8.12.  This value was 
computed from an optical study of the surrounding \ion{H}{2} region.  Also, we add the [O/H] value of 8.4 for NGC 1313,
determined separately by both \citet{cal94} and \citet{wal97}.

As shown in Figure~\ref{fig2}, our [O/H] values are consistently high compared to those obtained from the \citet{pil04} \ion{H}{2} study.  
\citet{pil04} include a discussion of how their values, obtained by the P-calibration method, are significantly
lower than those obtained by \citet{gar02} using the R$_{23}$-calibration method.  In Figure~\ref{fig2} we include
[O/H] values for NGC 253, NGC 300, M33, M81, and M83 from \citet{gar02} (triangles).  Our oxygen abundances are in much
better agreement with the values from this R$_{23}$-calibration method.

Our [O/H] values, which are consistent in the X-ray band between two separate absorption models,
({\tt tbabs*tbvarabs*edge*(grad + pow)} and an {\tt tbabs*edge*edge*(grad + pow)} model), are consistent with the values of \citet{gar02}.  We
further wish to compare them to the metallicity predicted from Sloan Digital Sky Survey (SDSS) results.
As a result of a SDSS study, \citet{tre04} found a luminosity-metallicity relation for their sample
of star-forming galaxies of: $12 + \log(O/H) = -0.185(\pm0.001)M_B + 5.238(\pm0.018)$.
In Figure~\ref{fig9} we compare our values with their results (represented by the line).  We obtained
absolute magnitudes (M$_B$) for the host galaxies using the total apparent corrected B-magnitude
recorded in the HyperLeda galaxy catalogue \citep{pat89}(parameter {\it btc}) and the distances listed
in Table 1 of \citetalias{win05}.  For M33 and NGC 4559, which were not included in the previous
study, we used the distances of 0.7\,Mpc and 9.7\,Mpc from \citet{ho97}.  The majority of our sources
are consistent with the SDSS results.  However, the sources in irregular galaxies have metallicities
much higher than predicted.   

\subsubsection{Galaxy Properties}
More luminous galaxies are sometimes expected to have higher star formation
rates, and thus higher metallicity.  However, we found no evidence of this.
Investigating further into the relationship between star formation rate (SFR) and metallicity, we chose
to look at a galactic luminosity diagnostic that is less dependent on extinction from dust, the 
infrared galactic luminosity (L$_{FIR}$).    In \citetalias{win05}, we calculated L$_{FIR}$ using data from the {\it Infrared Astronomical Satellite} and the approach
of \citet{swa04}.  We quoted values of L$_{FIR}$ in \citetalias{win05}.  Using the same method, with IRAS fluxes obtained from \citet{ho97},
we find M33 to have L$_{FIR} \approx 1.7 \times 10^{42}$\,erg\,s$^{-1}$ and NGC 4559 to have 
L$_{FIR} \approx 7.4 \times 10^{42}$\,erg\,s$^{-1}$.  We used these L$_{FIR}$ values to compare the SFR to both the oxygen and iron abundances.

\begin{figure}
\plotone{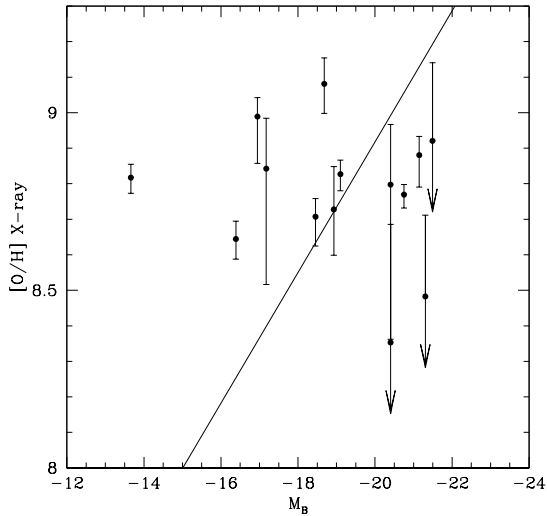}
\caption{Host galaxy M$_B$ vs. ULX X-ray [O/H] ratio.  The solid line represents the SDSS
results from a study of star-forming galaxies \citep{tre04}.  Arrows indicate that the lower limit
for the [O/H] X-ray parameter is below the plotted graph region.  Our values are largely consistent
with the Sloan results, with the exception of Holmberg IX and Holmberg II.  Both of these galaxies
are irregulars.  The X-ray [O/H] ratio derived from the ULX spectra is higher than predicted for these
objects. 
\label{fig9}}
\end{figure}

\begin{figure}
\plotone{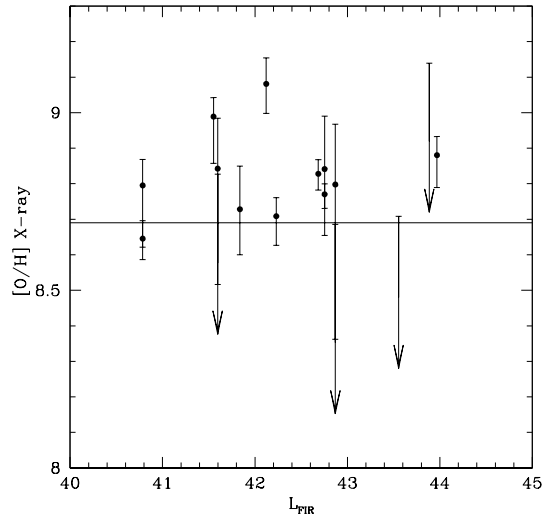}
\caption{Oxygen abundance ([O/H]) as a function of the host galaxy's FIR luminosity, obtained from 
IRAS.  The line represents the Wilms
solar value of [O/H] .  Arrows indicate that the lower limit
for the [O/H] X-ray parameter is below the plotted graph region.  There appears to be no correlation between L$_{FIR}$ and oxygen abundance.  In fact, the sources show
roughly solar [O/H] abundances regardless of host galaxy. 
\label{fig4}}
\end{figure}

\begin{figure}
\plotone{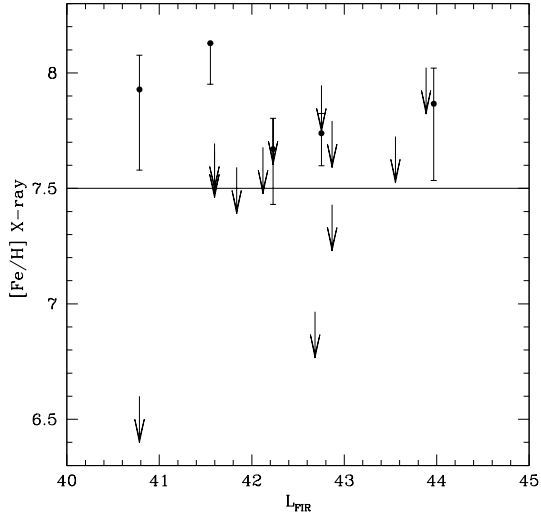}
\caption{Iron abundance ([Fe/H]) as a function of the host galaxy's FIR luminosity obtained from 
IRAS.  Holm IX did not have corresponding IR data, however, being an 
irregular galaxy, its L$_{FIR}$ would probably be comparable to Holm II.  The iron abundances of both of the irregulars
are well above those of the ULXs in spiral galaxies.  The sources in galaxies with larger L$_{FIR}$ have roughly equal 
iron abundances, within the error bars.  The iron abundances in all cases are well above the Wilms solar abundance.  Arrows indicate that the lower limit
for the [Fe/H] X-ray parameter is below the plotted graph region.
\label{fig5}}
\end{figure}

\begin{figure}
\plotone{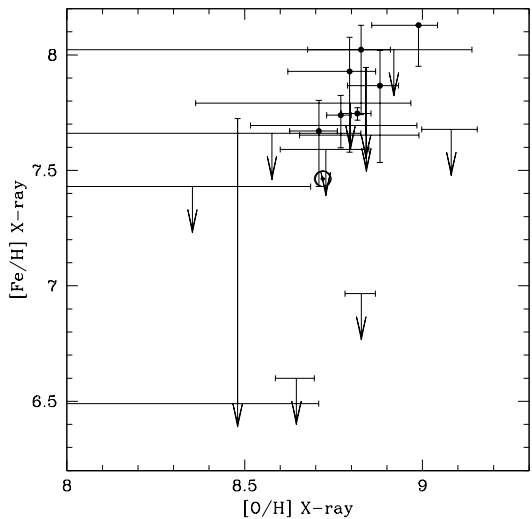}
\caption{[Fe/H] as a function of [O/H] from the {\tt tbvarabs} model, using the {\tt grad} and power law fit.  
The Wilms solar abundance is indicated with an
open circle.  The ratio of Fe/O abundances obtained through the X-ray spectral fits are 
approximately the Wilms solar values.  Arrows indicate that the lower limit
for the [Fe/H] X-ray parameter is below the plotted graph region.
\label{fig6}}
\end{figure}

\begin{figure}
\plotone{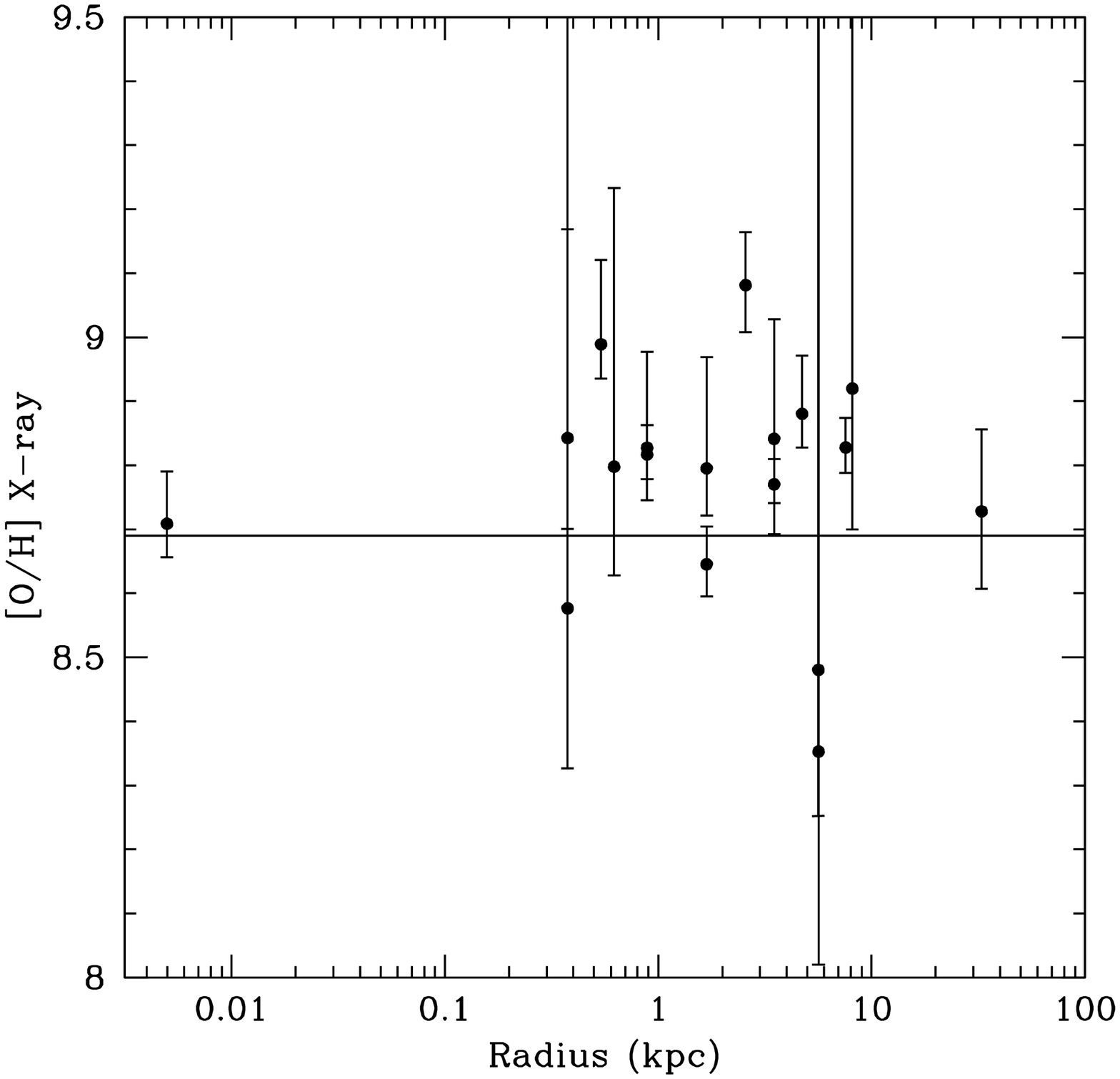}
\caption{Oxygen abundances from the {\tt tbvarabs} model vs. distance of the ULX from the host galaxy's dynamical center.  
The [O/H] ratios are given for spiral galaxies only.  There is no correlation between radius and oxygen abundance in our
sample.  However, since the sources originate in different host galaxies with different abundances, there is no conclusion
that can be drawn from this data.  The host galaxy NGC 4559, containing two ULX sources studied, does appear to show a
possible correlation (the source nearer the center has a higher oxygen abundance), however, the error bars on the
oxygen abundance are large for both sources. 
\label{fig3}}
\end{figure}


In Figure~\ref{fig4}, we plot oxygen abundance relative to SFR.  All sources with the exception of Holmberg IX XMM1,
which did not have available IRAS data, are plotted.   We see, as was also illustrated in Figure~\ref{fig9},
that the luminosity of the host galaxy does not determine the metallicity.  The more luminous galaxies
do not have metallicities higher than the less luminous galaxies.
In fact, most sources have oxygen abundances that are roughly the Wilms solar abundances (indicated
by the line).

In Figure~\ref{fig5} we see that there is no relationship between iron abundance (see
values in Tables~\ref{tbl-2} and ~\ref{tbl-3} for values)
and SFR.  The Wilms solar abundance for iron
is only $2.69 \times 10^{-5}$ relative the hydrogen abundance.  The plots show that the metallicity
relationship is very flat, all of the sources have roughly solar abundances.  This is also seen in
Figure ~\ref{fig6}.  This plot shows the [Fe/H] ratios
versus the [O/H] values, both obtained through the {\tt tbvarabs} model.  The solar Wilms
values are represented on the plot by the open circle symbol.  It appears that the sources
are slightly more abundant in iron than the solar value, however, the error bars are quite large.  The oxygen abundances, as before
stated, are roughly the solar Wilms value.  Roughly, the abundances appear solar.

With such a flat relationship between abundance and luminosity, we tested to see if this result carried
through in a comparison of abundance versus radial distance within the host galaxy.
This follows upon an interesting property observed in spiral galaxies. Namely, that abundances are typically higher in the center of
the galaxy and decrease with increasing
radius \citep{sea71}.  We tested how our results compare to this result in Figure ~\ref{fig3}.
For the sources located within spiral galaxies, we plotted the [O/H] ratio as a function of distance R from the
dynamical center (as reported by NED).  We used the distances quoted in Table 1 of \citetalias{win05} to translate angular
distance on the sky into kpc from the galactic center.   As evidenced in the plot, we
do not see much variation in the [O/H] values.  Clearly, the expected scaling of higher oxygen abundance towards
the center is not seen.  Since the sources lie in different host galaxies with different relative abundances,
it is possible that this trend may be detectable with a larger sample of galaxies.
However, the implication from Figures \ref{fig9} through \ref{fig3} is that the environment of the ULX
sources is relatively uniform in terms of metallicity.  The ULX sources appear to live in similar
environments, with metallicities roughly solar.

\section{Summary}
Through our work, we conclude that
X-ray spectral fits to ULX sources do provide a viable method of finding abundances in other galaxies. 
We have determined hydrogen column densities and oxygen abundances along the line of sight to 14 ULX sources.  To measure these values, we assumed a connection between ULXs and Galactic
black hole systems such that the ULX spectra used in this study correspond to a high/soft state.
Therefore, we modeled the sources with an absorbed disk model and power law (in XSPEC
{\tt tbabs} *{\tt tbvarabs}*({\it accretion disk model} + {\tt pow})).  We tested the effect different
accretion disk models have on the measurement of the host galaxy's hydrogen column density 
and elemental abundance with the disk models: {\tt grad}, {\tt diskpn}, {\tt diskbb}, and {\tt bbody}.
We found that the measured hydrogen column density and abundances are model independent.

We also tested the physical plausibility of this model by comparing the mass and mass accretion
rates obtained from the {\tt grad} model with expected results based on Galactic black hole
systems.  We found that the ULX spectra were consistent with the high/soft state, with 
$L/L_{Edd}$ values $< 1.0$ for
sources with a standard thermal model at low energies and power law dominated higher energy
spectrum.  For four sources, the spectra were consistent with a heavily Comptonized spectrum.
These sources are more likely stellar mass black hole systems in a very high state of accretion.
We modeled their spectra with a power law at low energy and disk model around 1\,keV.  This
model provided similar column density and abundance values to a more physical absorbed disk and
Comptonization model.

Comparing our X-ray measured column densities with those from optical and \ion{H}{1} studies for 
8 of our sources, we find that 7 of the sources have X-ray column densities 
approximately equal to those of the alternate methods.  This
implies that the hydrogen columns towards most of our ULX sources represent that of their host galaxy.
Since the \ion{H}{1} studies represent averages over a large beam area where the X-ray column densities are
directly along the line of sight to the ULX source, this implies that the ULX sources lie within
roughly normal areas of their host galaxies.  The exception in this study was M81 XMM1, whose X-ray
hydrogen column density was large relative to the \ion{H}{1} study.  This suggests that there is extra absorption
intrinsic to this source.     

The oxygen abundances appear to be  roughly the Wilms solar values.  For five sources,
the count rates were sufficient to determine iron abundances without large error bars
(see Figures~\ref{fig5} and ~\ref{fig6}).  
We found that iron abundances for these sources were slightly overabundant relative to 
the solar Wilms value.
However, within errorbars, the abundances appear solar.  X-ray derived [O/H] values are comparable
to those from an optical study by \citep{gar02}, indicating that the X-ray derived values are the same
as the [O/H] values of \ion{H}{2} regions within the host galaxy.
Luminosity-metallicity relationships
for the ULX host galaxies show a flat distribution, as does a radius-metallicity plot.  Therefore,
it appears that the ULX sources exist in similar environments within their host galaxy, despite the
wide range of host galaxy properties.

\acknowledgments
We would like to thank Rob Swaters and Robert Braun for help with the  \ion{H}{1} column densities and Tod Strohmayer for use of the 100\,ks Holmberg IX {\it XMM-Newton} data set before it became public.
We would also like to thank Tim Kallmann for useful discussion on the oxygen absorption edge and
an anonymous referee for suggestions which greatly improved this paper. 
This research has made use of the NASA/IPAC Extragalactic Database (NED) which is operated by the Jet Propulsion Laboratory, California Institute of Technology, under contract with the National Aeronautics and Space Administration as well as the HyperLeda database located online at:
\url{http://leda.univ-lyon1.fr/}.


\clearpage
\appendix
\section{Spectral Simulations}

We conducted spectral simulations in order to: (1) determine the number of counts needed
to measure the oxygen and iron abundances and (2) to verify the model independence 
of the galactic column density and abundances
with respect to the {\tt grad}, {\tt diskbb}, and {\tt bbody} models (seen in a comparison of
Tables~\ref{tbl-2} and~\ref{tbl-1}).  Towards this end, we created simulated pn spectra
based on the long Holmberg IX XMM1 observation's (0200980101) unbinned, pn spectrum.  We used
the base {\tt (grad + pow)} model parameters as indicated in Table~\ref{tbl-2}.  We modeled
both the Galactic column density (\citet{dic90} value: n$_H = 4.0\times10^{20}$\,cm$^{-2}$) and 
host galaxy column density (n$_H = 1.9\times10^{21}$\,cm$^{-2}$) with individual {\tt tbabs}
models.  Thus, all of the abundances were set to the solar Wilms values.  We used the
XSPEC command {\tt fakeit} to create simulated spectra with 200000, 40000, 10000, 
5000, and 2000\,counts.
The simulated spectra were binned with 20\,cts/bin using {\tt grppha}.  We fit the binned
simulated spectra with the models {\tt tbabs*tbvarabs*(grad + pow)}, {\tt tbabs*tbvarabs*(diskbb + pow)},
and {\tt tbabs*tbvarabs*(bbody + pow)}.  This allowed us to see the effects the different models
have on the measured galactic hydrogen column density and abundances.  The results for these
fits are seen in Table~\ref{tbl-7}.  The range that the oxygen and iron abundances were
allowed to vary within was 0.0 (lower limit) to 5.0 (upper limit) with respect to the solar values. 

\begin{figure*}
\plottwo{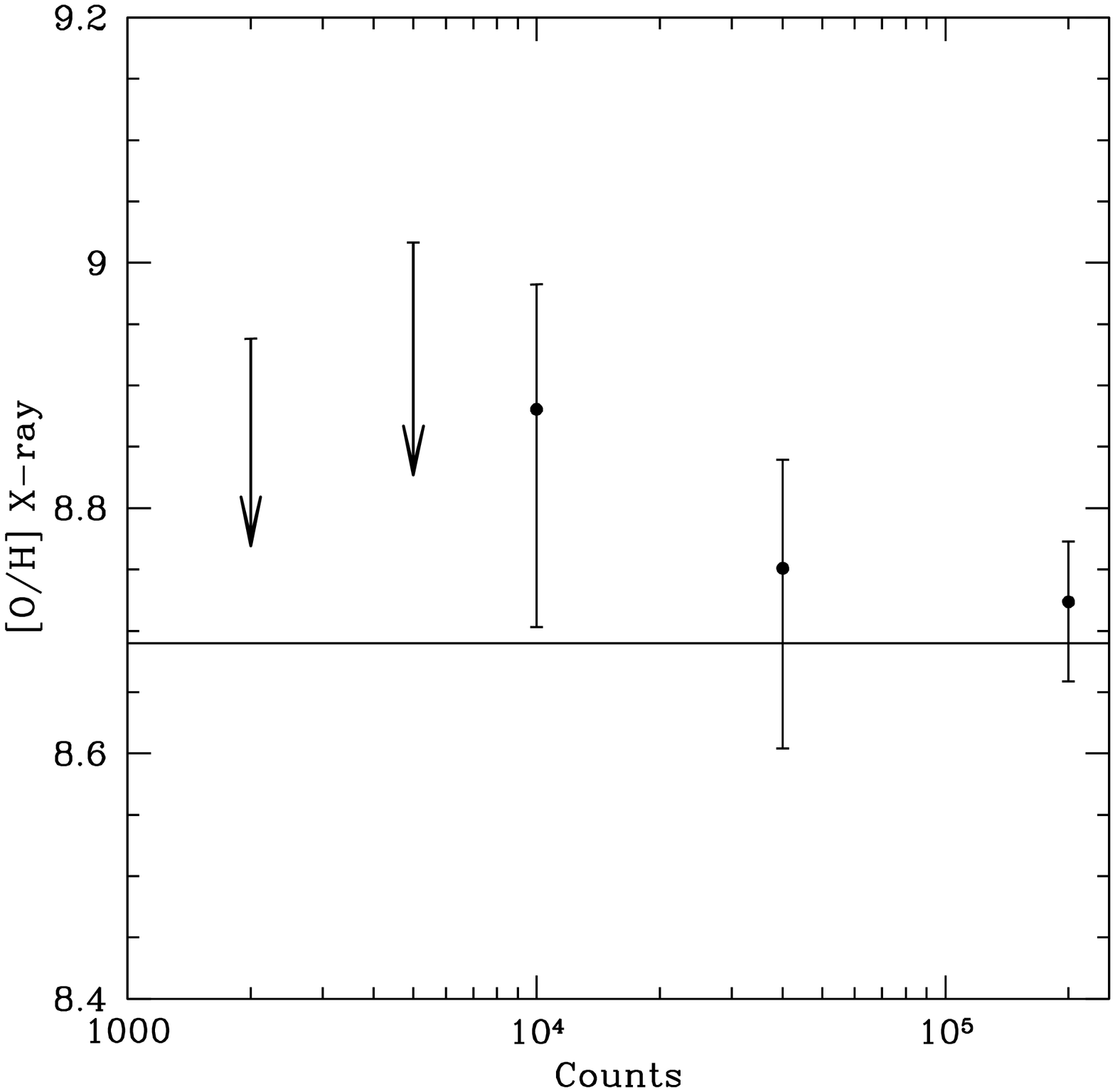}{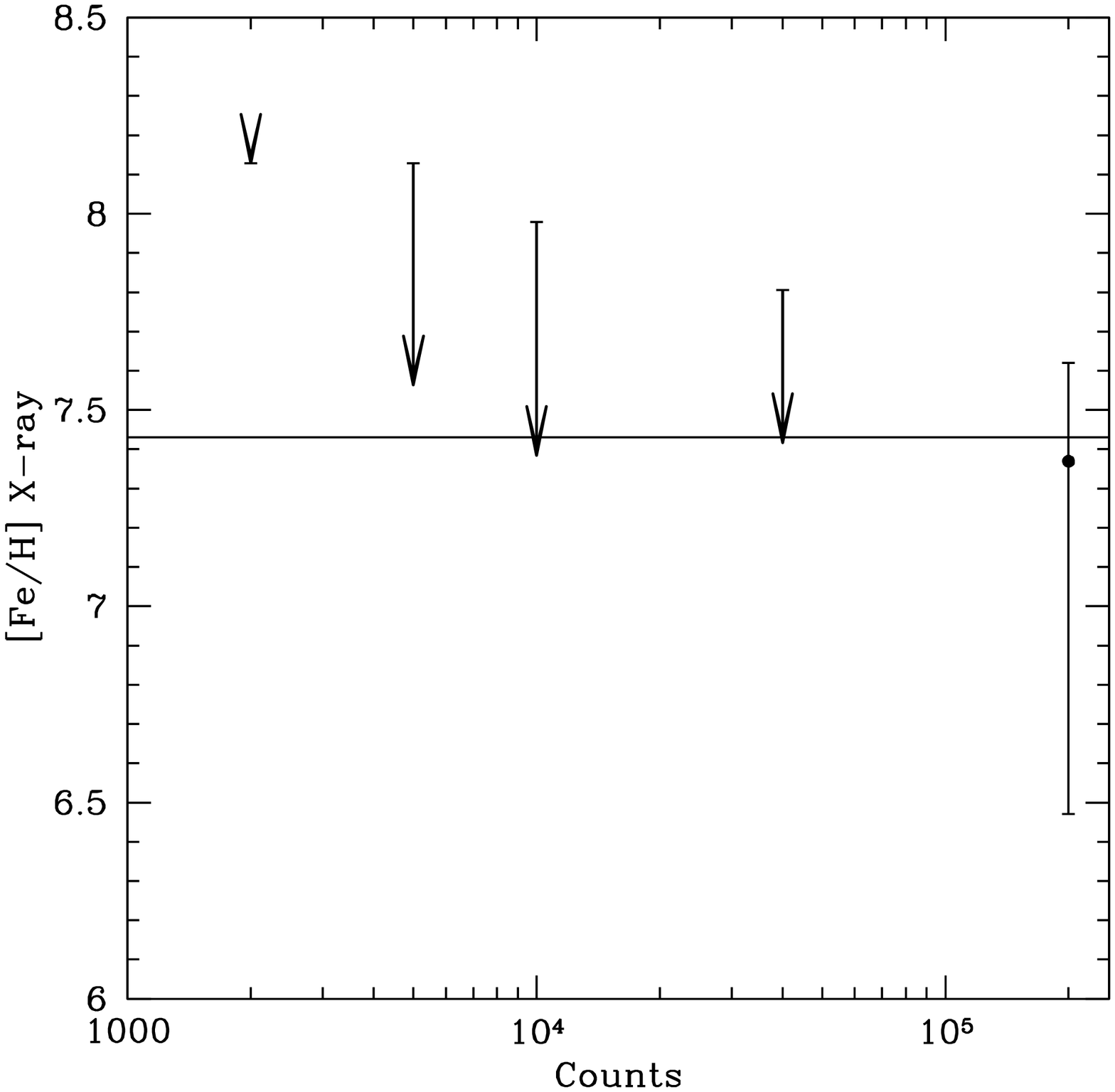}
\caption{The number of counts for simulated spectra versus the [O/H] value (left) and
[Fe/H] value (right) from the {\tt tbabs*tbvarabs*(grad + pow)}, where errors represent
the 90\% confidence rate.  The horizontal lines represent the solar Wilms values 
([O/H]$\approx 8.69$ and [Fe/H]$\approx 7.43$).  Arrows are used to represent errors that
extend below the range of the plot.  The tip of the arrow point represents the [O/H] or [Fe/H] 
parameter from the XSPEC model (see Table~\ref{tbl-7}).
\label{fig10}}
\end{figure*}

In Figure~\ref{fig10}, the number of simulated counts versus the errors on the oxygen 
and iron abundances are plotted for the {\tt tbabs*tbvarabs*(grad + pow)} model.  Here,
[O/H]$= 12 + \log(O\times0.00049)$ and [Fe/H]$= 12 + \log(Fe\times0.0000269)$, using
the solar Wilms values for O/H and Fe/H.  As
seen in the plots, the errors in oxygen abundance are much smaller for a given number of counts
compared to the errors in iron abundance.  Further, the upper limits on the oxygen abundance
continue to be meaningful through 2000\,counts.  This is not true for the iron abundances, where
the error bars extend through the entire range of allowed values (from Fe/H = 0.0 - 5.0, or
[Fe/H] up to 8.13).  Thus, our simulations show us that the iron abundance (from measurements
of the Fe L-shell edge at 851\,eV with the {\tt tbvarabs} model) requires at least 5000\,counts
to be detected.  At 5000\,counts the model derived value (Fe/H = 1.36) is meaningful, however, the
errors extend throughout the entire allowable range (Fe/H = 0.0-5.0).  The oxygen abundance
(from measurements of the O K-shell edge at 542\,eV with the {\tt tbvarabs} model) is detected
down to 2000\,counts, but with large errors below 10000\,counts.

From Table~\ref{tbl-7}, we find that the same trends described in Section 4.1 are present in
our simulations.  Namely, there is little variation between the model derived abundances
and column densities.  A comparison of the mean n$_H$ and oxygen
abundance values shows $\approx 2$\% difference between the {\tt grad} and {\tt diskbb} model.
The {\tt bbody} n$_H$ values are roughly 26\% higher while the oxygen abundances are
$\approx 11$\,\% higher.    Comparing the 40000 and 200000\,count spectra for the iron abundance,
we find that the {\tt grad} and {\tt diskbb} model values differ by $\approx 12$\% while the {\tt bbody}
model results are larger by a factor of 50\%.  While the {\tt bbody} model yields lower column
densities and higher abundances, the {\tt diskbb} and {\tt grad} models are in agreement.  The
differences in the {\tt bbody} results are low for the column density and oxygen abundance, but
appeared significant for the iron abundance.

\begin{deluxetable*}{lllll}
\tabletypesize{\scriptsize}
\tablecaption{Spectral Fits to Simulated HolmIX XMM1 {\tt tbabs*tbabs*(grad + pow)} Spectrum
\label{tbl-7}}
\tablewidth{0pt}
\tablehead{
\colhead{Counts\tablenotemark{a}} & 
\colhead{n$_{H}$\tablenotemark{b}} & \colhead{Oxygen abundance\tablenotemark{c}} & 
\colhead{Iron abundance\tablenotemark{c}} & \colhead{$\chi^{2}$/dof} }
\startdata
\cutinhead{{\tt tbabs*tbvarabs*(grad + pow)}}
200000 	& 0.21$^{+0.02}_{-0.02}$ & 1.08$^{+0.13}_{-0.15}$ & 0.87$^{+0.68}_{-0.76}$ & 1687/1727 \\
\\
40000	& 0.19$^{+0.04}_{-0.03}$ & 1.15$^{+0.26}_{-0.33}$ & 0.97$^{+1.41}_{-0.97}$ & 933/914 \\
\\
10000	& 0.20$^{+0.07}_{-0.06}$ & 1.55$^{+0.41}_{-0.52}$ & 0.90$^{+2.64}_{-0.90}$ & 340.5/401 \\
\\
5000	& 0.16$^{+0.13}_{-0.07}$ & 1.37$^{+0.75}_{-1.34}$ & 1.36$^{+3.64}_{-1.36}$ & 187.2/201 \\
\\
2000	& 0.33$^{+0.14}_{-0.19}$ & 1.20$^{+0.57}_{-1.20}$ & 5.0$^{+0.0}_{-5.0}$ & 71.9/89 \\
\cutinhead{{\tt tbabs*tbvarabs*(diskbb + pow)}}
200000 	& 0.21$^{+0.02}_{-0.02}$ & 1.09$^{+0.15}_{-0.19}$ & 0.99$^{+0.94}_{-0.76}$ & 1691/1727 \\
\\
40000	& 0.19$^{+0.03}_{-0.03}$ & 1.16$^{+0.27}_{-0.35}$ & 1.09$^{+1.47}_{-1.09}$ & 934/914 \\
\\
10000	& 0.19$^{+0.03}_{-0.05}$ & 1.58$^{+0.43}_{-0.50}$ & 1.01$^{+2.80}_{-1.01}$ &  341/401\\
\\
5000	& 0.16$^{+0.12}_{-0.08}$ & 1.41$^{+0.76}_{-1.16}$ & 1.57$^{+3.43}_{-1.57}$ & 187.5/201 \\
\\
2000	& 0.32$^{+0.18}_{-0.19}$ & 1.23$^{+0.59}_{-1.23}$ & 5.0$^{+0.0}_{-5.0}$ & 71.7/89 \\
\cutinhead{{\tt tbabs*tbvarabs*(bbody + pow)}}
200000 	& 0.17$^{+0.02}_{-0.02}$ & 1.24$^{+0.16}_{-0.20}$ & 1.65$^{+0.85}_{-1.01}$ & 1702/1727 \\
\\
40000	& 0.15$^{+0.04}_{-0.02}$ & 1.28$^{+0.36}_{-0.40}$ & 1.73$^{+1.97}_{-1.72}$ & 937/914 \\
\\
10000	& 0.15$^{+0.07}_{-0.04}$ & 1.77$^{+0.61}_{-0.72}$ & 1.22$^{+3.78}_{-1.22}$ & 343.7/401 \\
\\
5000	& 0.13$^{+0.10}_{-0.07}$ & 1.61$^{+0.97}_{-1.61}$ & 2.19$^{+2.81}_{-2.19}$ & 187.5/201 \\
\\
2000	& 0.21$^{+0.19}_{-0.12}$ & 1.16$^{+0.92}_{-1.16}$ & 5.0$^{+0.0}_{-5.0}$ & 71.8/89 \\
\enddata
\tablenotetext{a}{Total number of photon counts for simulated pn spectrum.  Simulated spectra
created with the XSPEC {\tt fakeit} command, using the model parameters and
response files from the long
HolmIX XMM1 observation.}
\tablenotetext{b}{Hydrogen column density determined from {\tt tbvarabs} in units of $10^{22}$\,cm$^{-2}$.  The Galactic value of n$_H$ was fixed to the \citet{dic90} value with the {\tt tbabs} model.}
\tablenotetext{c}{Element abundance relative to the Wilms solar abundance from the {\tt tbvarabs} model}

\end{deluxetable*}

\clearpage
\section{Comparison with SAS 6.5 data processing}
We reprocessed the pn data in this study with the new version of SAS (6.5), in order to observe
the effects different SAS versions would have on our results.  After processing the data (with
{\tt epchain}), we extracted spectra as described in Section 2.  The spectra were then fit with
the {\tt tbabs*tbvarabs*(grad + pow)} model.  Note that the {\tt edge} model was not needed to correct
for differences in calibration between the MOS and pn since we only fit pn spectra.  Since the MOS data
was not added, the counts for the pn only spectra are lower than the MOS + pn spectra.  
Thus, we excluded spectra for sources with
less than 5000\,counts in the pn.  The results are shown in Table~\ref{tbl-8}, ordered by number of counts.

\begin{figure*}
\plottwo{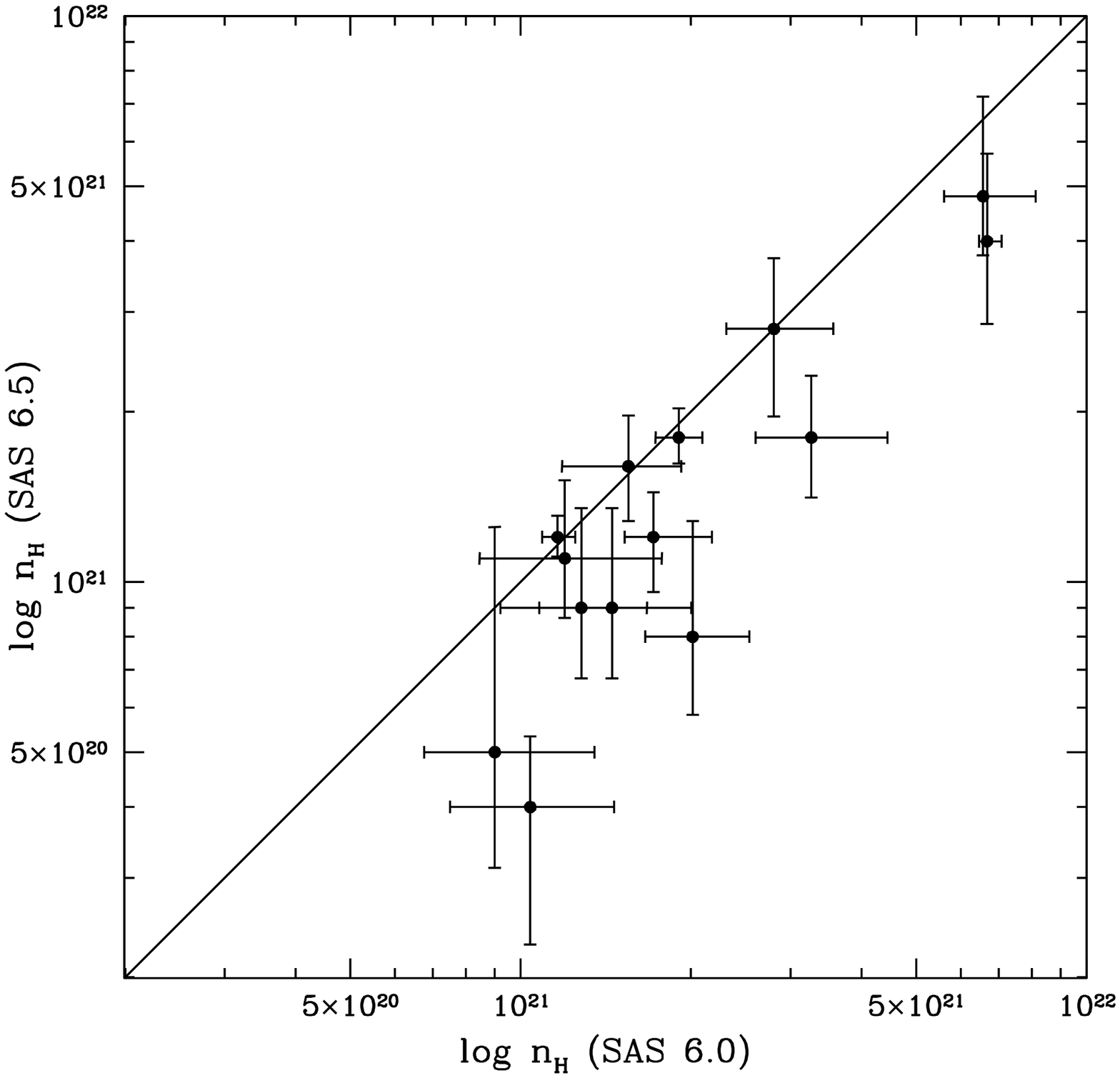}{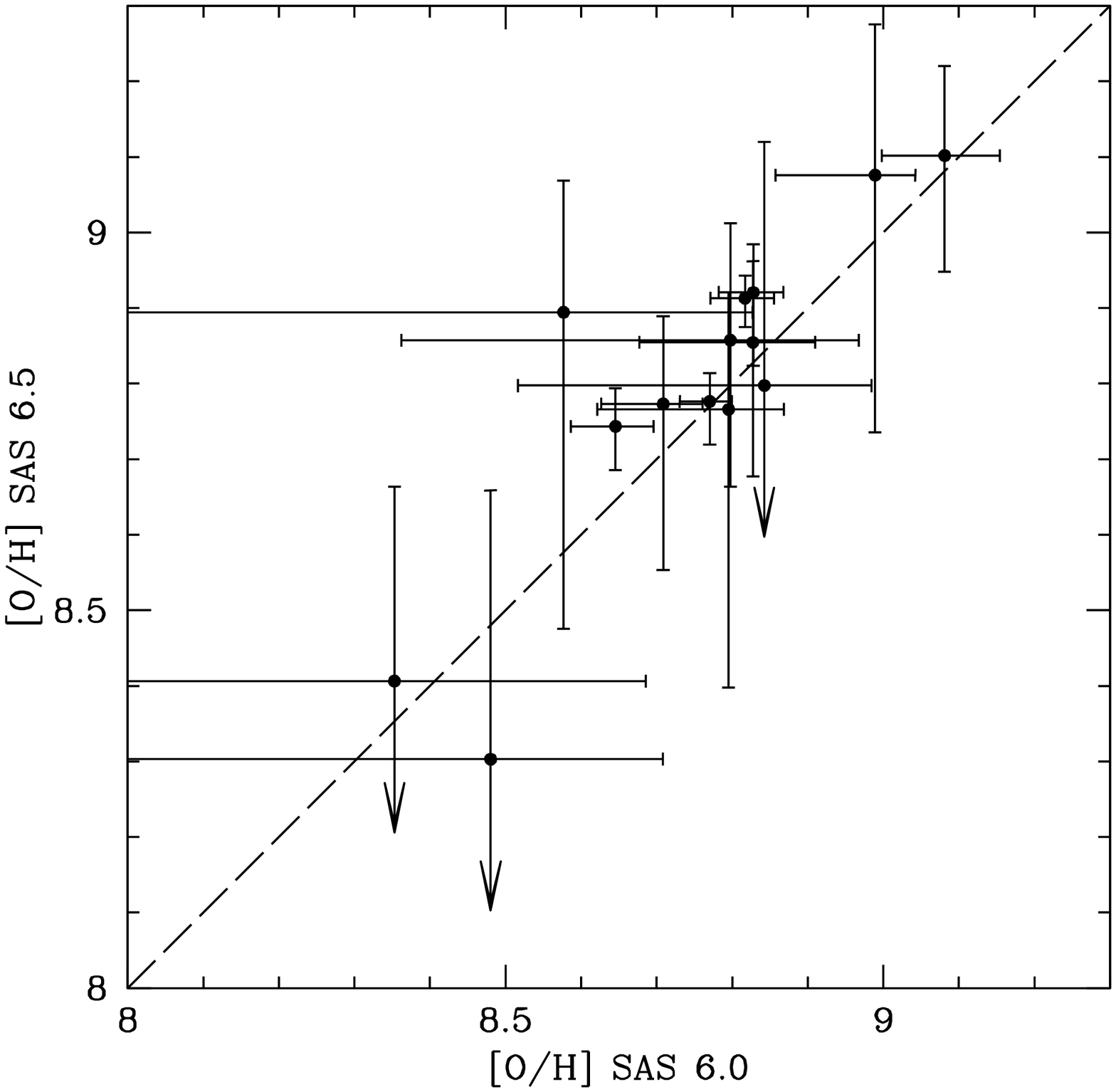}
\caption{Comparison of the galactic/ULX hydrogen column density obtained from the {\tt tbvarabs} model
for SAS 6.0 processed data (from Table~\ref{tbl-2}) versus SAS 6.5 processed data 
(from Table~\ref{tbl-8}) (left) and comparison of the
similarly obtained [O/H] values (right).  From these plots, we find that the SAS 6.0 n$_H$ and [O/H]
values are consistent with the SAS 6.5 values.  The version 6.0 column densities are slightly lower
than the 6.5 values and the version 6.0 oxygen abundances are slightly higher, on average.
The lines in the plot represent agreement between the SAS 6.0 and 6.5 values.  Arrows are
used to indicate fits where the lower limit for the oxygen abundance extends below the
plotted range. 
\label{fig-sas}}
\end{figure*}

In Figure~\ref{fig-sas}, we plot a comparison of the X-ray hydrogen column densities and oxygen 
abundances
(both from the {\tt tbvarabs} model) from SAS version 6.0 and 6.5.  From
these plots, we find that the SAS 6.0 n$_H$ values are slightly lower but consistent with the
SAS 6.5 observations.  The [O/H] values are slightly higher for SAS 6.0, but are also consistent.
Since the hydrogen column density and oxygen abundance values are consistent between both
versions, we did not find it necessary to recalculate EPIC spectral parameters for the sources.
Thus, the values quoted in the paper are from the data processed with SAS version 6.0.  
While the error bars for the SAS 6.5 Fe abundances are larger, due to the lower number of
counts in the pn alone, within the errors the values are consistent with the SAS 6.0 values.  
For the highest number of count sources (Holmberg IX XMM1 and Holmberg II XMM1), the
fits for the longest observations yielded the same Fe abundance for Holmberg II XMM1 
(Fe/H = 0.0 ranging to $\approx 0.20$) and consistent values for Holmberg IX XMM1 (the SAS 6.0 
values
(Fe/H from 2.19 to 1.94) were well within the SAS 6.5 values (Fe/H from 2.37 to 0.75)).

We also
note that the release notes for SAS version 6.0 (Carlos Gabriel \& Eduardo Ojero, available on-line)
indicate no substantial differences in the EPIC responses.  Therefore, the
{\tt edge} model correction quoted in \citet{bau05} is adequate and did not need to be recalculated from SAS 5.4 (as in \citet{bau05}) to SAS 6.0.

\begin{deluxetable*}{lllllllll}
\tabletypesize{\scriptsize}
\tablecaption{Spectral Fits for EPIC spectra with {\tt tbabs*tbvarabs*(grad + pow)} model\label{tbl-8}}
\tablewidth{0pt}
\tablehead{
\colhead{Source} & \colhead{n$_{H}$\tablenotemark{a}} & \colhead{O abund.\tablenotemark{b}} 
& \colhead{Fe abund.\tablenotemark{b}} & \colhead{Mass (M$_{\sun}$)} 
& \colhead{$\dot{M}/\dot{M}_{Edd}$\tablenotemark{c}} & \colhead{$\Gamma$} & \colhead{$\chi^{2}$/dof} 
& \colhead{counts\tablenotemark{d}}
}
\startdata
HolmIX XMM1& 0.18$^{+0.02}_{-0.02}$ & 1.67$^{+0.12}_{-0.14}$ & 1.61$^{+0.76}_{-0.88}$ & 442$^{+110}_{-91}$ & 0.08$^{+0.02}_{-0.02}$ & 1.47$^{+0.02}_{-0.02}$ & 1637.1/1532 & 171210\\
\\
\nodata& 0.18$^{+0.05}_{-0.04}$ & 1.46$^{+0.41}_{-0.49}$ & 1.84$^{+2.23}_{-1.84}$ & 258$^{+327}_{-164}$ & 0.10$^{+0.12}_{-0.06}$ & 1.68$^{+0.08}_{-0.09}$ & 477.8/486 & 14996\\
\\
HolmII XMM1& 0.12$^{+0.01}_{-0.01}$ & 1.13$^{+0.14}_{-0.14}$ & 0.0$^{+0.24}_{-0.0}$ &  100$^{+26}_{-23}$ & 0.11$^{+0.03}_{-0.03}$ & 2.45$^{+0.06}_{-0.05}$ & 939.3/932 & 150290\\
\\
\nodata& 0.12$^{+0.03}_{-0.02}$ & 1.19$^{+0.51}_{-0.68}$ & 3.08$^{+1.92}_{-3.08}$ & 272$^{+232}_{-163}$ & 0.07$^{+0.06}_{-0.03}$ & 2.31$^{+0.12}_{-0.13}$ & 362.4/368 & 13983\\
\\
M33 X-8& 0.08$^{+0.03}_{-0.03}$ & 1.21$^{+0.37}_{-0.48}$ & 0.0$^{+1.15}_{-0.0}$ & 7.23$^{+0.68}_{-0.63}$ & 0.96$^{+0.06}_{-0.06}$ & 2.12$^{+0.25}_{-0.26}$ & 714.9/749 & 46579\\
\\
M81 XMM1& 0.48$^{+0.13}_{-0.16}$ & 1.22$^{+0.11}_{-0.15}$ & 1.61$^{+0.58}_{-1.09}$ & 9.13$^{+0.73}_{-0.83}$ & 3.29$^{+0.18}_{-0.31}$ & 3.31$^{+0.68}_{-0.95}$ & 865/811 & 42188\\
\\
N4559 X-7 & 0.16$^{+0.04}_{-0.03}$ & 0.52$^{+0.42}_{-0.52}$ & 0.0$^{+1.08}_{-0.0}$ & 1000$^{+1057}_{-604}$ & 0.04$^{+0.05}_{-0.03}$ & 2.16$^{+0.07}_{-0.08}$ & 407.1/359 & 12492\\
\\
NGC5204 XMM1& 0.04$^{+0.03}_{-0.01}$ & 1.28$^{+1.41}_{-1.28}$ & 0.0$^{+5.0}_{-0.0}$ & 120$^{+146}_{-80}$ & 0.04$^{+0.05}_{-0.02}$ & 1.94$^{+0.10}_{-0.13}$ & 305.7/323 & 9991\\
\\
\nodata & 0.09$^{+0.03}_{-0.03}$ & 1.60$^{+0.79}_{-0.99}$ & 0.0$^{+2.43}_{-0.0}$ & 254$^{+194}_{-104}$ & 0.10$^{+0.04}_{-0.04}$ & 1.90$^{+0.22}_{-0.25}$ & 244.8/278 & 7503\\
\\
NGC4559 X-10 & 0.11$^{+0.03}_{-0.03}$ & 1.47$^{+0.63}_{-0.53}$ & 0.0$^{+2.0}_{-0.0}$ & 7.06$^{+3.45}_{-3.0}$ & 4.76$^{+2.65}_{-2.83}$ & 2.22$^{+0.58}_{-0.29}$ & 266.4/328 & 8834\\
\\
NGC300 XMM1 & 0.09$^{+0.03}_{-0.03}$ & 2.58$^{+0.81}_{-0.77}$ & 0.0$^{+1.67}_{-0.0}$ & 160$^{+153}_{-81}$ & 0.01$^{+0.01}_{-0.01}$ & 2.46$^{+0.15}_{-0.16}$ & 278.9/237 & 6771\\
\\
NGC1313 XMM3 & 0.40$^{+0.16}_{-0.12}$ & 1.70$^{+0.27}_{-0.34}$ & 3.0$^{+1.46}_{-1.77}$ & 907$^{+1477}_{-695}$ & 0.06$^{+0.14}_{-0.01}$ & 2.25$^{+0.09}_{-0.17}$ & 246.2/232 & 6237\\
\\
NGC5408 XMM1 & 0.05$^{+0.03}_{-0.03}$ & 2.43$^{+1.42}_{-1.32}$ & 5.0$^{+0.0}_{-5.0}$ & 840$^{+361}_{-245}$ & 0.11$^{+0.04}_{-0.03}$ & 2.24$^{+0.31}_{-0.30}$ & 174/187 & 5928\\
\\
NGC4631 XMM1 & 0.28$^{+0.12}_{-0.07}$ & 0.41$^{+0.52}_{-0.41}$ & 0.0$^{+2.5}_{-0.0}$ & 6.73$^{+2290}_{-6.4}$ & 0.72$^{+1.15}_{-0.72}$ & 1.95$^{+0.24}_{-0.40}$ & 215.4/189 & 4830\\
\enddata
\tablenotetext{a}{Hydrogen column density determined from {\tt tbvarabs} in units of $10^{22}$\,cm$^{-2}$.  The Galactic value of n$_H$ was fixed to the \citet{dic90} value with the {\tt tbabs} model.}
\tablenotetext{b}{Element abundance relative to the Wilms solar abundance from the {\tt tbvarabs} model}
\tablenotetext{c}{Ratio of mass accretion rate from the {\tt grad} model to Eddington accretion rate (see Section 4)}
\tablenotetext{d}{Total number of photon counts from the pn detector}
\end{deluxetable*}

\end{document}